\documentclass[12pt,preprint]{aastex}

\shorttitle{Radio, X-ray Observations of the Lk Ha101 Cluster}
\shortauthors{Osten \& Wolk}

\newcommand{\e}{et al.\ }

\newcommand{\minusone}{$^{-1}$}

\newcommand{\II}{\protect\small II \normalsize $\!\!$}

\newcommand{\nh}{\mbox{$N_{\rm H}$}}

\newcommand{\skipthis}[1]{}

\newcommand{\asec}{$^{\prime\prime}$}

\newcommand{\psqcm}{{\rm cm}^{-2}}

\newcommand{\ps}{{\rm s}^{-1}}

\newcommand{\erg}{{\rm ergs}}

\newcommand{\be}{\begin{equation}}
\newcommand{\ee}{\end{equation}}

\newcommand{\amin}{$^{\prime}$}

\begin{document}

\title{Multiwavelength Signatures of Magnetic Activity from Young Stellar Objects in the LkH$\alpha$101 Cluster}
\author{Rachel A. Osten\altaffilmark{1}\altaffilmark{2}}
\affil{Astronomy Department, University of Maryland, College Park, MD 20720 U.S.A.}
\email{osten@stsci.edu}
\author{Scott J. Wolk}
\affil{Harvard-Smithsonian Center for Astrophysics, 60 Garden St., Cambridge, MA 02138 U.S.A.}
\email{swolk@cfa.harvard.edu}

\altaffiltext{1}{Hubble Fellow}
\altaffiltext{2}{Currently at Space Telescope Science Institute, 3700 San Martin Drive, Baltimore, MD 21218}

\begin{abstract}
We describe the results of our multi-wavelength observing campaign
on the young stellar objects in the LkH$\alpha$101 cluster.
Our simultaneous X-ray and multi-frequency radio observations are unique in providing 
simultaneous constraints on short-timescale variability at both wavelengths
as well as constraints on the thermal or nonthermal
nature of radio emission from young stars.  Focussing in on radio-emitting objects
and the multi-wavelength data obtained for them, we find that
multi-frequency radio data indicate nonthermal emission even in objects with
infrared evidence for disks.
We find radio variability on timescales of decades, days and hours. 
About half of the objects with X-ray and radio detections were variable
at X-ray wavelengths, despite lacking large-scale flares or large variations.
Variability appears to be a bigger factor affecting radio emission than X-ray emission. 
A star with infrared evidence for a disk, [BW88]~3, was observed in the decay phase of radio flare.
In this object and another ([BW88]~1),
we find an inverse correlation between radio flux and spectral index which
contrasts with behavior seen in the Sun and active stars.
We interpret this behavior as the repopulation of the hardest energy electrons
due to particle acceleration.
A radio and X-ray source lacking an infrared counterpart, [BW88]~1,
may be near the substellar limit; its radio properties are similar to
other cluster members, but its much higher radio to X-ray luminosity ratio is reminiscent of
behavior in nearby very low mass stars/brown dwarfs.
We find no correspondence between signatures
of particle acceleration and those of plasma heating, both time-averaged and time-variable.
The lack of correlated temporal variability in multi-wavelength behavior, the 
breakdown of multi-wavelength correlations of time-averaged luminosities, and
the optical thickness of X-ray-emitting material at radio wavelengths support the
idea that radio and X-ray emission on young stars are physically and/or energetically 
distinct.
\end{abstract}


\section{Introduction}
Magnetic fields play an important, but poorly understood, role in the formation and further
evolution of late-type stars.
The influence of magnetic fields in stellar environments can be seen, for example, in X-ray emission
from gas which has been magnetically heated to coronal temperatures,
and radio emission arising from the action of nonthermal
particles in the presence of magnetic fields.
In the last 20 years, remarkable advances in the detection and
study of stellar sources at X-ray and radio wavelengths have revealed 
an intimate connection between radio and X-ray luminosities of 
late-type stars.  
This relationship extends from time-averaged
estimates of the presumably steady emission, to spectacular 
examples of flare enhancements when studied simultaneously with
multiple wavelengths.
Yet the underlying connection is still
not well constrained. 
Most of the observational multi-wavelength studies
of magnetically-induced flare activity on late-type stars have 
concentrated on targeted studies of nearby, extremely active stars.
The goal of the current project is to investigate
a region of active star formation, using simultaneous X-ray and radio  
observations, to constrain the relationship of these emissions in both 
steady and varying emission. This will allow us to  
determine the extent to which the solar paradigm can be
carried in these very active young
stars.

\subsection{The Stellar X-ray--Radio Connection}
At first glance, there is an obvious disconnect between X-ray and radio emission
from active stars:  X-ray observations are well described by optically thin,
thermal, high temperature plasma generally seen in a distribution between
a few MK and tens of MK (few keV), while radio observations generally probe nonthermal
populations of
accelerated electrons with energies of a few MeV.  
Yet among
several classes of active stars and solar flares, a tight relation 
between X-ray and radio luminosities is
found, L$_{X}/$L$_{R}$ $\sim 10^{15.5 \pm 0.5}$ Hz \citep{gb1993,bg1994}. 
This relation extends from solar microflares up to the most energetic
coronal sources (in radio and X-rays): the weak T Tauri stars (wTTs; Class III objects).  The observational
relationship invites an interpretation based on a common origin -- particle acceleration
and thermal coronal heating, separate but related
steps in the flow of energy after a magnetic reconnection event,
as inferred from solar and stellar flare studies.


On the Sun, the observed relationship between nonthermal processes and thermal coronal
emission during individual transient impulsive flare events is well known \citep{neupert1968}
and has given rise to a standard flare scenario.
Particles are
accelerated high in the corona in connection with a reconnection event,
and downwardly-directed electron beams ``inject'' these energetic particles
into coronal loops, which act as a magnetic trap.  Radio gyrosynchrotron
emission occurs from trapped particles; those which precipitate 
due to pitch-angle scattering encounter the dense chromosphere
where they are collisionally stopped, depositing their energy in the 
ambient plasma.
The temperate chromospheric
material becomes heated to coronal temperatures 
on timescales short compared with the
hydrodynamic expansion timescale, and undergoes a radiative instability,
expanding up into the corona, where the high temperature gas radiates at X-ray
wavelengths. 
The X-ray radiation increases gradually during the impulsive
particle acceleration episode, reaching a peak after the acceleration has
ended and subsequently declining to previous values \citep{fisher1985}.
Radio gyrosynchrotron emission
from the accelerated electrons is roughly proportional to the
injection rate of electrons, whereas the X-ray luminosity is roughly
proportional to the accumulated energy in the hot plasma.  This simple
scenario gives
rise to the 
observed L$_R \propto dL_X/dt$ seen in the Sun; as the amount of
nonthermal energy input (diagnosed by the radio emission) increases,
the thermal radiative output (X-ray emission) also increases, until the acceleration and
subsequent heating ceases and
the system settles back into quiescence.
Observations of some stellar flares at
multiple wavelengths confirm this scenario for a variety of active stars
\citep{gudel1996,hawley2003,gudel2002,osten2004,smith2005},
yet there is also ample evidence that the situation on stars is more complex.

\subsection{Radio Emission from Young Stars}
Either thermal or nonthermal radio emission from young stellar objects can be
expected to be present.  
Thermal sources are more likely to be identified with disks, or with
higher mass objects with winds \citep{skinner1993}; nonthermal
emission is associated with particle acceleration and magnetic reconnection.  
Based on the classification sequence of young stellar objects from infalling
protostar to wTTs and the decreasing relative importance of accretion and
disk material versus magnetic activity \citep[see review by][]{fm1999}, we expect that Class 0, I and II 
(a.k.a classical T Tauri stars; cTTs) sources 
will be more dominantly thermal radio sources, while Class III sources
are potential nonthermal radio sources.  
There are observations which suggest
that nonthermal radio emission has been detected in a few Class I
objects \citep{feigelson1998},
raising the possibility that nonthermal diagnostics of magnetic activity can be
produced in Class I and II objects, but obscured by free-free absorption in the 
accretion disk or wind.
These two cases can be discerned with
multi-frequency radio observations, as the spectral shape of the continuum emission will be different.
Nonthermal emission can additionally be discerned due to short time-scale variability 
from magnetic reconnection processes; circular polarization from
structures containing large-scale magnetic fields provides an additional observational
constraint.
In addition to gyrosynchrotron emission in
T Tauri stars,
evidence exists also for coherent emission 
\citep[highly
circularly polarized emission;][]{smith2003}. 

The X-ray and radio emission from wTTs are
enhanced orders of magnitude above the Sun and most 
active stars' variability extremes,
signaling a possible extension of the trends from the lower 
activity stars, or a transition
to a new kind of phenomenon. 
T Tauri stars show similar behaviors to the well-studied 
active stars (and the Sun itself) ---
starspots, luminous X-ray and radio emission, 
enhanced chromospheric emission
\citep{fm1999}, signalling the importance of magnetic fields
in producing these signatures in young stars as well. 
Magnetic fields have been measured on the surfaces of
young stars \citep{cmj2007}; while their generation in
wTTs is probably not due to an interface dynamo as in the Sun
due to the fully or nearly fully convective nature of
wTTs, alternative explanations such as convective or turbulent dynamos (described 
in \citet{ck2006} and \citet{dobler2006})
may be responsible for magnetic field production in wTTs. 
It is not known to what extent magnetic
activity phenomena are affected by circumstellar
material and/or jets and ionized winds, as can be
found around young low-mass stars.
While some wTTs do show correlations between non-simultaneously obtained time-averaged X-ray and radio
luminosities \citep{gb1993,bg1994}, other studies have
found little or no correlation between these luminosities or their
variability \citep{gagne2004}.  
Confirmation of the complex nature of
magnetic activity signatures also arises 
from the lack of multi-wavelength correlations between X-ray and radio emission
in Class~I sources 
\citep{forbrich2006,forbrich2007}.

Radio observations are an inefficient method of finding cluster objects, due
to the low radio detection rates: 
radio surveys find radio emission rates of $<$10\% for CTTs \citep[generally objects associated with optical jets and
Herbig Haro objects,][]{bieging1984}, 
$\sim$20\% for Herbig Ae/Be stars \citep[due to thermal emission from a wind;][]{skinner1993}, and 
10--50\% for WTTs \citep[emission due to magnetic activity;][]{white1992}.
The asymmetry in multi-wavelength associations for young stars was 
pointed out by 
\citet{andre1987}, who noted from their study of the $\rho$ Oph
core that a majority of stellar radio sources are
detected at X-ray and IR wavelengths, but the converse is not true.
\citet{gagne2004} found similar results 
in the $\rho$ Ophiuchi cloud; in their study, 10 sources were detected at both
radio and X-ray wavelengths, compared with 31 radio sources and
87 X-ray sources. 
Biases from distance, sensivity, and variability play a role in these trends.
Despite the low radio detection rate, radio observations 
are important in the context of magnetic activity signatures in that
they providue unique diagnostics of accelerated electrons unavailable at other wavelengths,
and deep observations can reveal new sources heretofore undetected in 
shallow radio surveys.

The purview of radio variability studies on T Tauri stars have usually been
long timescales of months to years,
due to the survey nature of most of the radio observations.  
There is some evidence
that similar features are seen in radio flares from wTTs as on hyperactive
stars:  short timescale (hrs) flux enhancements \citep{feigelson1994}, spectral index increases
during flux enhancements signalling switch from optically thin to
optically thick emission \citep{felli1993}, and detection of circular polarization \citep{white1992}.




\subsection{The Target --- The LkH$\alpha$101 Cluster}
LkH$\alpha$ 101 is a luminous ($\sim 5\times10^3L_\odot$) 
Herbig Be star with a strong wind \citep{barsony1990},
an associated H\II~ 
region (Sharpless-222) and a reflection nebula (NGC 1579). 
The visual extinction in 
the extended area is about 1 magnitude.  IRAS data show 
100$\mu$m emission extending 30$^\prime$ around the star.
Recent observations by \citet{tuthill2001,tuthill2002}
have discovered that this star possesses a disk
which is nearly face on, and shows evidence of a
secondary star.  
This cluster was originally identified by the ``necklace of radio sources'' 
around the central star LkH$\alpha$101 \citep[][hereafter, BW88]{bw},
as they
detected 9 point sources at 6 cm using the VLA.  
Subsequent observations by \citet{so} (hereafter, SO98)
revealed an additional 16 radio sources. 
\citet{ab1994}
found 51 sources (K$<$16.8) within 40$^{\prime\prime}$ of LkH$\alpha$101.
Extinction of these sources is moderate (3-20 A$_V$) and  
one-third of the stars show
infrared excesses consistent with disks.  The 2MASS data 
clearly show a small cluster of stars, centered on LkH$\alpha$101,
about 3$^\prime$ in radius (the extent of the optical nebula). 
 Comparison of this field with nearby
2MASS pencil beams indicates about 65 more sources in this direction
within the 6$^\prime$ diameter.  This is not the whole
cluster as  \citet{ab1994} found sources 2 magnitudes fainter than
the 2MASS limit. 
We take the distance to the cluster to be 700 pc \citep{herbig2004};
a detailed discussion of distance estimates occurs in
a more comprehensive paper on the wealth of X-ray, infrared, and 
optical data on the cluster (Wolk et al., in prep.).

In this paper, we discuss simultaneous {\it Chandra} ACIS-I and VLA observations of
a young PMS cluster to bridge the gap between the well-studied
X-ray--radio connection of the nearby active stars and 
that of the more distant and more common (and potentially more
magnetically active) pre-main sequence stars.  
Such studies of high energy processes in star forming regions have the
advantage of numerous stars in the field of view to increase the ``stellar monitoring time''
and thereby increase the likelihood of observing transient emissions.
The focus on this paper is the radio observations and 
the insights into magnetic activity in young stars which the simultaneous 
radio and X-ray observations provide.
Section 2 describes the data reduction, \S 3 describes the analysis,
\S 4 presents and discusses the results in terms of the magnetic structures 
giving rise to radio and X-ray emissions, \S 5 discusses the implications of
the results, and \S 6 concludes. 

\section{Data Reduction}
The primary goal of this program was the simultaneous $Chandra$ and
VLA observations to investigate variability in active young stars.  
To achieve the desired sensitivity
in the X-ray bands we required 80ks (22hr) total observation time.  In order
to enable nearly completely simultaneous coverage of the $Chandra$
observations with the VLA, we divided the 80 ks into two 40~ks (11 hr)
observations separated by 2 days, occurring on 6 and 8 March, 2005.
We also drew on available
non-contemporaneous 2MASS and $Spitzer$ data to help elucidate the
nature of the sources in the field.  A companion paper (Wolk et al.,
in prep., hereafter Paper~II) presents a detailed analysis of the Chandra and {\it Spitzer}
data and accompanying optical data including 2MASS, Spitzer IRAC
as well as MIPS 24 $\mu$m data.  For the present purposes we use
the derived X-ray and IR source positions, IR photometry, and X-ray variability
and spectral shape to construct a better picture of the radio sources.
In this section, we outline the
observations and basic reduction of the radio and X-ray data.

\subsection{Radio Data Reduction}
The LkH$\alpha$101 cluster was observed by the VLA\footnote{The National Radio Astronomy Observatory is a facility of the National Science Foundation operated under cooperative agreement by Associated Universities, Inc.}
in March 2005 in project S60872.  The observing setup consisted of splitting the array into two sub-arrays, and
observing simultaneously at 3.6 and 6 cm.  Observations were obtained in this fashion on 6 and 8
March, for observing sessions spanning 10.5 and 11 hours,
respectively.
The approximate full width at half maximum of the primary beam at 3.6 and 6 cm is
5.3 and 9.4 arcminutes, respectively, meaning that the longer wavelength has sensitivity
extending over larger distances from the cluster center.  
Thus, while there was only one pointing direction, limiting the field of view to a fraction of
the Chandra field, the deep exposure combined with simultaneous multi-frequency
recording ensures an ability to constrain the nature of short time-scale variability
for radio-emitting sources within $\sim$5\amin\ of LkH$\alpha$101.
The phase center of the observations, or ``aimpoint'', was 2\asec\ south of LkH$\alpha$101.
The array was in B configuration, affording a reasonable compromise
between spatial resolution and wide-field coverage; as the bright central Be star is the source
of a radio-bright H~II region there is a fair amount of extended emission.
Radio data was reduced and calibrated in AIPS; the primary flux calibrator was 3C48
and the phase calibrator was 0414+343.  After the initial calibration, phase self-calibration
of the cluster was done using the bright central source.  
Because of the large amount of
extended emission from the nebula, visible at small $uv$ distances, 
images were made with restricted baseline lengths; after examining the behavior of the correlated 
flux versus baseline, X-band (3.6 cm) maps were made with $uv$ distances restricted
to 30--300 k$\lambda$, C-band (6 cm) maps were made with $uv$ distances restricted to 10--200 k$\lambda$.
Even with these $uv$-range restrictions, there appear to be radio-emitting structures
larger than the fringe spacing of the shortest baselines; dark bowls around the central object
and large-scale stripes in the image (particularly at 6 cm) indicate this.
Images at each frequency were made
using visibilities for each day as well as combined for both days; thus 6 maps of the region around
LkH$\alpha$101 were generated and searched for radio sources.
The maps were corrected for the primary beam by dividing by the primary beam gain factor.
C-band maps were made using pixel sizes of 0.3\asec$/$pixel, X-band maps made with 0.15\asec$/$pixel,
with image sizes 2048 pixels, corresponding to 5.1\amin\  and 10.2\amin\  for X and C-bands, respectively, thus
covering the FWHM of the primary beam.
However, the non-monochromatic nature of the radio
emission results in the intensity of a point source being reduced at larger distances
from the pointing center; the total (integrated) 
flux is preserved but the peak flux decreases in proportion to
the bandwidth relative to central frequency, and the source offset in units of the synthesized
beam.  Because source detection is based on peak flux density exceeding 
a threshhold, the effect of bandwidth smearing will degrade our attempts to find weak point sources
far from the center of the cluster. 

In addition to creating images of total intensity at each wavelength, we also imaged the same
regions in Stokes V to search for any significant amount of circularly polarized flux, as
has been noted from previous observations of young stellar objects \citep{white1992}.
We did not detect any statistically significant levels of circularly polarized flux.
The Stokes V image rms at 3.6 cm was 10 $\mu$Jy, and the Stokes V image rms at 6 cm was 13$\mu$Jy.

We generated a catalog of previously detected radio sources in this cluster, using the
source lists of \citet{bw} and \cite{so}, precessed to J2000 coordinates.
We follow the SIMBAD naming convention for these sources based on their designators in these
two papers.
In addition we searched for new sources in the combined maps at each frequency.  
All radio sources were given names in accordance with IAU convention,
and have the acronym ``LkHa101VLA'' accompanied by a sequence giving the J2000
coordinates, although we retain the use of the earlier names in previously
identified objects.
Based on our initial visual inspection of the combined images,
six new sources were identified using a minimum signal to noise (peak/rms) of 5.
We added these positions to the catalog of previously identified sources, and performed
a systematic determination of fluxes at each frequency and observing day: fits were done to
an 11 $\times$ 11 pixel square around the source coordinates when converted from RA, Dec to image pixel location.
For previously identified sources, we recorded a detection if the signal to noise ratio exceeded 3.
Fluxes were determined by two-dimensional Gaussian fits to the primary beam-corrected images.  Correction
for bandwidth smearing was done.  For previously detected sources, we determined the flux or upper limit
at the position of the source as noted in other papers.  
Some previously identified sources were not within the field of view 
of our observations.  Table~\ref{tbl:radiotbl1} lists the 3.6 and 6 cm fluxes obtained from 
co-adding all the data from this epoch,
as well as previously reported flux densities, taken from the literature \citep{bw,so}.

\subsection{X-ray Data Reduction}

The field was observed by $Chandra$ on 6 March 2005 starting at 17:16
UT  for 40.2 ks of total time and 39.6 ks of so-called ``good-time'' 
(ObsId 5429). It was observed again on 8 March 2005 starting at 17:43 UT
for essentially the same duration (ObsId 5428). The ACIS was
used in the nominal imaging array (chips I0-I3) which provides  
a field of view of approximately 17\arcmin $\times$ 17\arcmin.
The aimpoint was at 04:30:14.4, +35:16:22.2  (J2000.0) with a roll of 281 degrees.

The data 
were processed through the standard CIAO pipeline at the
Chandra X-ray Center, using their software version DS7.6.  
For the purposes of point source detection, the data from the two
observations were merged into a single event list following
established CIAO procedures to create a merged event list. 
To identify point sources, photons
with energies below 300 eV and above 8.0 keV were filtered out from
this merged event list.  
A monochromatic exposure map was generated in the standard way using an
energy of 1.49 keV which is a reasonable match to the expected 
peak energy of the sources and the $Chandra$ mirror transmission.  
WavDetect was then run on a series of flux
corrected images binned by 1, 2 and 4 pixels. The resulting
source lists were combined and this resulted in the detection of
about 210 sources.
Detailed description of X-ray data reduction, and analysis,
including point source extraction and flux estimation, is given in Paper~II. 

 
%
%


\section{Analysis}

\subsection{Multi-Wavelength Source Identification \label{sec:rxrir}}
\begin{figure}
\includegraphics[scale=1.0]{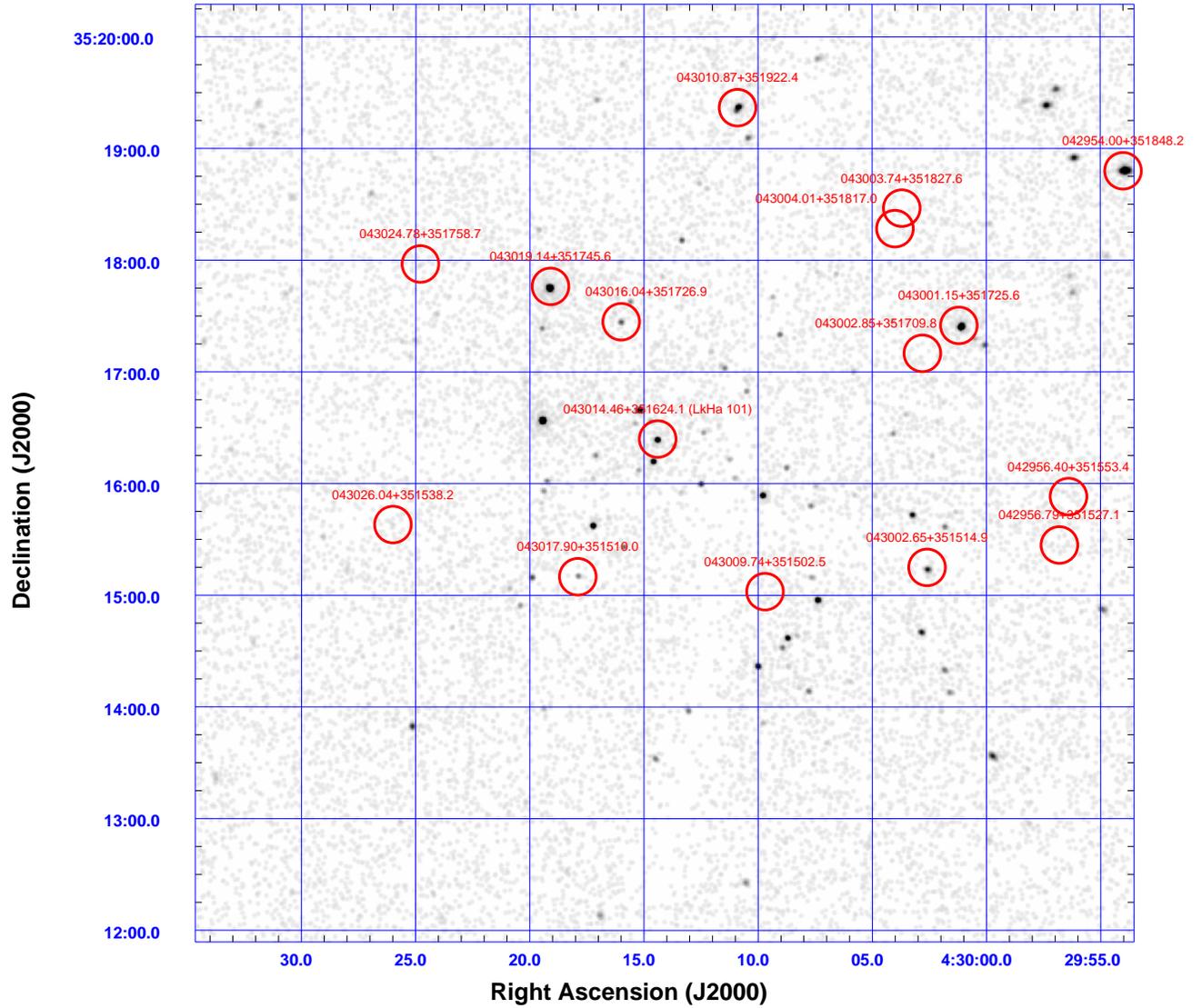}
\caption{ Close-up of the portion of the X-ray image also covered by
  the VLA field of view.  The locations of the 16 radio sources considered in this paper 
are indicated with red circles. 
Radio sources have the prefix LkHa101VLA J prepended to the positions.
\label{fig:XrayVLAfield}}
\end{figure}

The VLA source positions were searched against the IRAC point source catalog (from Paper~II)
for matches.
An offset of 1\asec\ produced a total of 7 matches; relaxing the required
positional coincidence to as much as 4 \asec\ did not produce any additional matches.
Of these, 4 are Class~III objects, 2 are Class~II objects, and
one (the central star LkH$\alpha$ 101) is classed as unknown.
The VLA source positions were also cross-correlated against the Chandra
source positions, with a positional coincidence of 1\asec\ allowed for
X-ray sources within 200\asec\ of the pointing center, and 2.5\asec\ allowed for
X-ray sources greater than 200\asec\ from the pointing center. 
Within 5\amin\ of LkH$\alpha$101, there are 132 X-ray sources, with 
100 having counterparts in the IR catalog, 2 being
Class~I, 35 being Class~II, 40 being Class~III, and
23 being of unknown type.  
Figure~\ref{fig:XrayVLAfield} shows the portion of the Chandra image
covered by the VLA field, with positions of radio sources indicated.
Using the analytic expressions in \citet{gehrels1986} for the
1 $\sigma$ confidence limits for small numbers of events in
Poisson statistics, the rate of radio emission for X-ray-emitting Class~II
objects is 2/35 or 6$^{+7}_{-4}$\%, and for
Class~III objects  it is 4/40 or 10$^{+8}_{-5}$\%.
We also searched the tables of \citet{herbig2004} for correspondences with
radio sources.
The identification of the radio sources at other wavelengths is given in
Table~\ref{tbl:idtbl}.

The radio observations reported on the sources around LkH$\alpha$101
span $\sim$20 years
with the VLA, taken at different frequencies and configurations.  
Because several sources have reported flux densities at only one epoch, and because we
expect variability to be a key characteristic of some of these sources, we extracted the
previous epochs from the VLA archive, reduced, calibrated and imaged the data, 
and fitted Gaussians at the reported positions of
all sources.  We reproduced three out of the four epochs reported in BW88, as the fourth
epoch (reported to have occurred on 1986 May 12) was not in NRAO's on-line archive.  We found different
flux densities for [BW88] 7 and 8 on 1986 April 25: our fitted flux density for source 7 was
consistent with that reported in BW88 for source 8, and vice versa.  This may be the result of typographical
error. 
We could not recover the new sources which were described in SO98 with our evaluation of the archival data.
Although fluxes at $>$3$\sigma$ uncertainty were determined at these positions in other epochs,
the unrecoverability of the sources along with absence of supporting data at other wavelengths
led us to exclude the SO98 sources from consideration in subsequent analyses.
Table~\ref{tbl:radiotbl1} lists the flux densities for sources as reported in BW88, SO98, and
by combining the data from our two days of observations.

The estimated number of foreground X-ray sources is small, and should not contaminate
the sample of X-ray and radio coincidences.
Our limiting X-ray sensitivity is about 10$^{-15}~\erg~\psqcm~\ps$ in the 0.5-2.0 keV band (\S 3.5). 
The results from the Champlane \citep{hong2005} survey for $Chandra$
fields in the plane of the Galaxy {\it but} away from the galactic core
indicate that we can expect to find about 70 background AGN, cataclysmic
variables, neutron stars, black holes, and other non--PMS star
point sources in a sample of 15 ACIS observations in the galactic
plane, but not pointed at the
galactic center, $(90^{\rm o} < l < 270^{\rm o})$, sensitive to similar flux levels as we have.
The number is most likely lower than this due
to the nearly opaque nature of the dust clouds in the central $\sim$40\% of the field, perhaps
45 total, or only $\sim$4 X-ray non-PMS point sources
scaled to the 5 \amin overlap between the $Chandra$ field and the VLA 3.6 cm field, or
a contamination of only 3\% of the X-ray sources.
Many of these are among the faint X-ray sources,
X-ray sources with no optical/IR counterparts or the sources with offset between the
X-ray position and optical/IR position exceeding 4\asec. 

It is unlikely that the radio sources lacking
X-ray and IR/optical counterparts are foreground sources.
As discussed in \S 3.2, most of these objects have negative radio spectral indices,
indicating nonthermal emission.  
Galactic sources of nonthermal radio emission 
should also produce nonthermal X-ray emission at a level above our detection limit (see \S 3.5);
the lack of X-ray detection implies a ratio of X-ray to radio flux less than $\approx$6$\times$10$^{11}$ Hz.
The lack of infrared or optical counterparts also constrains the presence of possible thermal
radio sources; as discussed in the following paragraph, the Spitzer maps are sensitive to
cluster members near the substellar mass limit, and foreground stellar objects will be brighter.
In order to produce thermal emission at radio wavelengths but faint optical light
a distance modulus that would place the objects beyond the cluster (and therefore not a
foreground object) would be needed.
Additionally, the space density of galactic foreground objects capable of being radio-bright but
optical or X-ray faint is small enough that within the 5\amin\ radius of LkH$\alpha$101
we do not expect to see any of these sources.

We estimate the upper limit on IR magnitude 
for an object not to be detected in the IRAC bands, using the
background levels due to nebulosity at the positions of the
radio sources.  We searched for the faintest star within 36\asec\
of the position of each radio source not having an IR detection, and
grouped sources with similar amounts of background.
We find typical limits in JHK of 17, 15.5, and 15.3, respectively.
At a  distance modulus of 9.225,  
a $K_{mag}$ of 15.3 corresponds roughly to an absolute K magnitude of $\sim$6.  
According to \citet{siess2000} a 1MY star with 0.1 M$_{\odot}$
would have a K$_{\rm abs}$=4.39.  DUSTY models 
\citep{chabrier2000,baraffe2002}
predict K$_{\rm abs}$=6 for a 0.06 M$_{\odot}$
star, near the stellar-substellar boundary.
Thus, the lack of an IR counterpart to apparent cluster members which are
detected at other wavelengths
may indicate that the object is a very low mass star or brown dwarf.

Eight of the radio sources considered here remain without a counterpart at X-ray or IR wavelengths.
The reliability of these sources can be examined by the number of
epochs or frequences at which each source has been detected. 
These eight sources have each been detected at more than one frequency or epoch;
one source (LkHa101 J043004.0+351817)
was identified in our observations at a single frequency on both days, as well
as being marginally detected in the archival data from 1985 April (see Table~\ref{tbl:radiotbl1}).
All of these sources are detected at $>$5 $\sigma$.
Several of these are possible extragalactic sources.
\citet{bw} identified their source [BW88]~5 as an extragalactic double.
A check of the NRAO VLA Sky Survey \citep{nvss}\footnote{Catalog available at 
http://www.cv.nrao.edu/nvss/} reveals three sources near LkH$\alpha$ 101 which are coincident
(to less than 10'') with sources [BW88]~7, [BW88]~5, and [BW88]~2.  As the integrated flux densities of these NVSS
sources at 20 cm are 5.5, 4.1, and 3 mJy, respectively, which are all higher than the 
range of flux densities reported at the higher frequencies considered here, 
it is likely that these are extragalactic
sources with steep radio spectral indices. 
The lack of coincidence with X-ray or IR sources
supports this conjecture.
Using a model for the 1.4 GHz extragalactic source counts, and converting from
1.4 GHz to 8.4 GHz using a spectral index of $\alpha$=-0.7, 5$_{-2}^{+3}$ sources are
expected in a circle of diameter 5.3' with flux density greater than the 3 $\sigma$ rms at 8.4 GHz of 
36 $\mu$Jy (J. Condon, private communication).  This is consistent with all 8 radio sources 
lacking a counterpart at other wavelengths being extragalactic in nature.

\subsection{Radio Spectral Index Measurements \label{alpha}}
The parameter which describes the shape of continuum radio emission is the spectral index $\alpha$,
defined as $S_{\nu} \propto \nu^{\alpha}$; generally, for $\alpha \ge$ 0.5 the spectrum
is rising, for $\alpha \le -0.5$ the spectrum is falling, and in between ($-0.5 \le \alpha \le 0.5$)
the spectrum is considered flat.  
Rising spectra can generally be associated with optically thick thermal emission, falling spectra with
optically thin nonthermal emission. Flat spectra could, in principle, be associated with either, but 
usually indicate nonthermal emission from an inhomogeneous source.
Most of the sources for which spectral index constraints are available indicate negative
spectral indices and consequently a nonthermal interpretation.
Table~\ref{tbl:radiotbl1} lists the spectral index derived from maps made by combining
the two days of radio observations
reported here; 13 objects have constraints on spectral index from detection at one or more frequencies.
Eight of the 13 have spectral indices which are flat and slightly 
negative ($-0.8\le \alpha \le -0.1$); 
two are steeply negative ($\alpha \le -1$);
another two have large positive spectral indexes, one of which is
the central star to the cluster, LkH$\alpha$101, and the other is a newly discovered source.
Based on the multi-wavelength source identifications discussed in \S~\ref{sec:rxrir},
four objects with IR or X-ray detections have spectral index constraints, and 3/4 of these
have negative spectral indices.  Out of three objects with IR detections and
spectral index constraints, two have negative spectral indices, one being a Class~II object
and the other a Class~III object.
Table~\ref{tbl:radiotbl2} lists the spectral index measurements on each day.  These are
generally consistent with the values determined from the combined maps.
The discrepant sources appear to have varied more at one wavelength than at the other.
This behavior is discussed more in section~\ref{sec:avar}. 

\subsection{Radio Variability}

\subsubsection{Long Timescale Radio Variability \label{longt}}
In order to
assess long-term radio variability, we considered the five epochs of 6 cm observations.
We determined an object's variability over the timescale of $\sim$ 20 years by examining the range of
recorded flux values (including epochs where only upper limits were obtained) and
determining how many $\sigma$ this range represents, using the largest measurement error.  Flux 
determination can be affected by the beam size (if the object is slightly resolved), distance from
the phase center, and contribution of background nebulosity.  Since the observations were made in
different array configurations, we place a conservative limit of 10$\sigma$ on the range of
maximum to minimum flux values to assess variability.  
Using this criterion, 4 out of 9 sources can be considered variable;
for objects detected on 4 or more occasions, only two can be considered variable.
Table~\ref{tbl:varstat} lists the range of maximum to minimum variations in units of $\sigma$
for the 9 objects detected on one or more occasions at 6 cm.

\subsubsection{Two-Day Radio Variability \label{mediumt}}
We examined variations of the radio sources 
for changes between the two days of our recent observations.
Table~\ref{tbl:radiotbl2} lists the fluxes and spectral indices determined for each day and frequency.
For all objects with at least one detection in our four intra-epoch maps (2 frequencies x 2 days), we
examined the flux densities and error bars of each day to determination if there was
statistical evidence for variability between the two days; we considered a source variable
at one frequency if the two flux density measurements differed by more than 3$\sigma$, where
the $\sigma$ of both flux density measurements was used.  If an object was undetected on one day
but detected on the other, it was considered variable if the detected flux was more than 3$\sigma$
above the upper limit.  These results are recorded in Table~\ref{tbl:varstat}.
Using these criteria, 4 objects are considered variable at X-band, and 6 objects are considered
variable at C-band.

\subsubsection{Short Timescale Radio Variability \label{shortt}}
\begin{figure}
\includegraphics[scale=0.7]{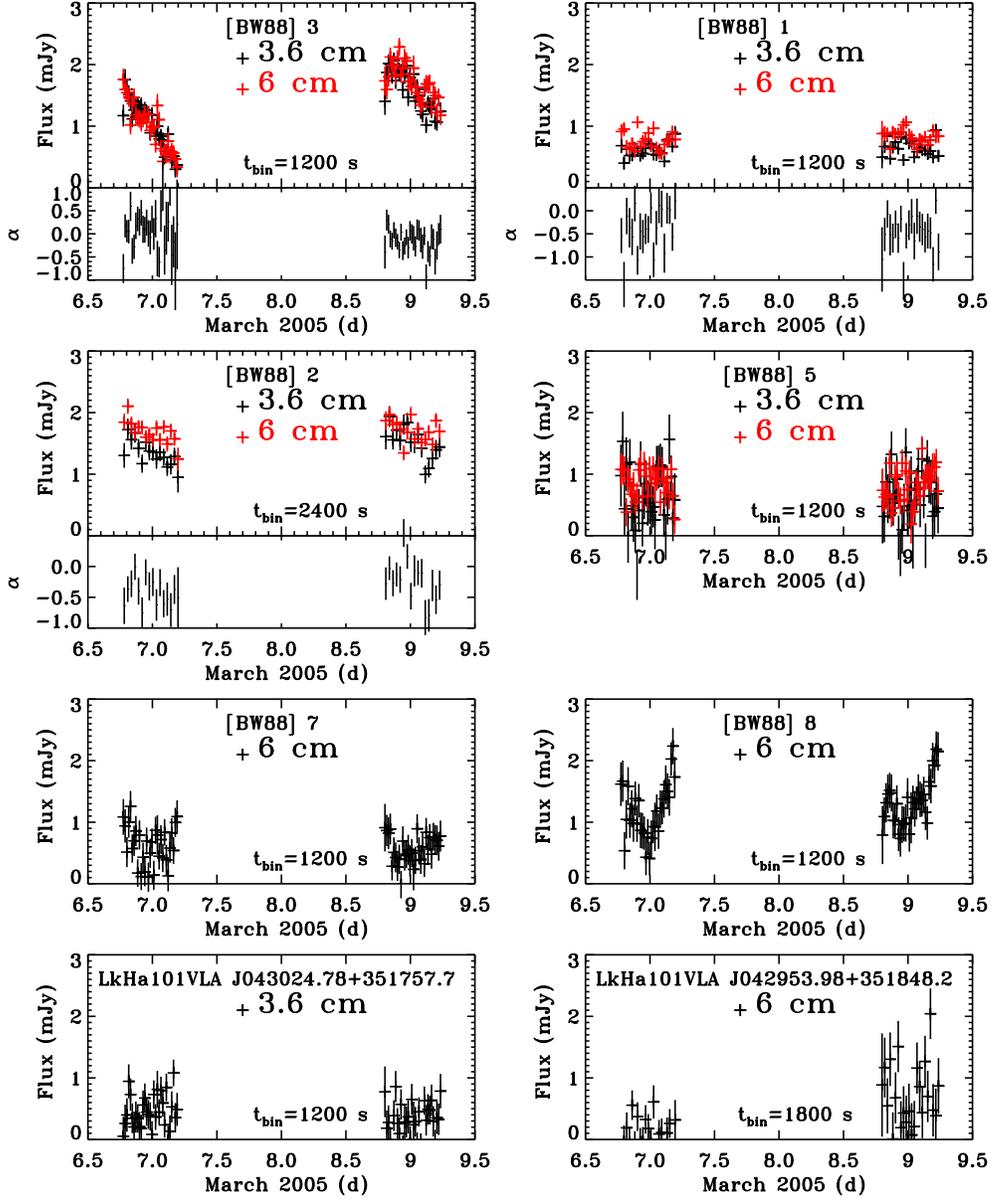}
\caption{Short-term variability amongst the strongest radio sources.  For objects where multi-frequency
information is available and the individual flux measurements are strong enough, the time variation
of the spectral index is plotted. 
\label{fig:radiolc}}
\end{figure}

For radio sources with strong detections ($>10\sigma$) in the combined maps, 
we explored variability on time scales shorter
than the 11 hour integrations each day.
In order to do this, we subtracted the visibilities of all other sources from the
calibrated visibility dataset, producing a dataset containing visibilities from only the target of interest.
We imaged the result to ensure that no additional flux contribution from the other
sources was present.  
We then searched for time variations in the visibility data using the AIPS task
$DFTPL$, with offset fixed to the position of the source, and $uv$ distances 
restricted to the values used in imaging.
We do not include in this analysis the radio-bright central object LkH$\alpha$ 101,
due to its immersion in the extended H~II emission region as well as 
the central ionized wind \citep[imaged at higher frequencies
in ][]{gibbhoare2007}, which makes it difficult to remove
the visibilities of the extended emission.  Also, based on its large positive spectral index,
indicative of a thermal wind source, we do not expect to see variations in the flux density
on such timescales.
The daily averaged flux densities listed in Table~\ref{tbl:radiotbl2} were measured 
from an image which had been corrected for the effects of the primary beam, while the
visibility datasets used to measure short-timescale flux density variations do not include the
effect of the primary beam.  In order to compare the daily-averaged flux densities with the
variations on shorter time-scales, we scaled the light curve flux densities 
by the primary-beam-corrected
flux densities so that the average flux density from the light curve on each day matched
the value in Table~\ref{tbl:radiotbl2}.
Four sources were strong enough at both 3.6 and 6 cm to enable this; additionally, there were two
sources which
were only
detected at 6 cm due to the larger field size, and one source 
with a stronger detection at 3.6 cm than at 6 cm.
A final source was only detected on one day at one frequency.
This selection resulted in 12 light curves, which are displayed in Figure~\ref{fig:radiolc}.
We made trial light curves using a number of different sized time bins, to explore evidence for variability
given signal-to-noise constraints.  

In order to asses evidence for variability from these light curves, we computed the average
and standard deviation of the light curve flux values, determining the $\chi^{2}_{\nu}$ statistic
and associated probability
for variations about the average value.  We also computed the distribution of fluxes and compared
that with a Gaussian distribution with the same mean and standard deviation as the data.
We computed the $\chi^{2}_{\nu}$ value between these two distributions
as another measure of how the data depart from an expected constant value with scatter.
Table~\ref{tbl:varstat} lists the probability that the observed $\chi^{2}_{\nu}$ statistic
could be exceeded by a random variable for each of these two methods of assessing variability.  We
considered there to be evidence for short-term variability if both of these methods had
probabilities of 5\% or less.  This resulted in two out of five objects showing evidence 
for variability at 3.6 cm, and four out of seven objects at 6 cm.


In addition,we also searched for statistically significant amounts of circularly polarized flux which
might appear transiently during any flare activity, but not be statistically significant
in the combined image.  Our examination of the variation of Stokes V flux at the position of
each of the target sources did not reveal any evidence for transient amounts of Stokes V flux. 

\subsection{Radio Variability and Spectral Index Changes\label{sec:avar}}
\begin{figure}
\begin{center}
\hspace*{-1cm}
\includegraphics[scale=0.4]{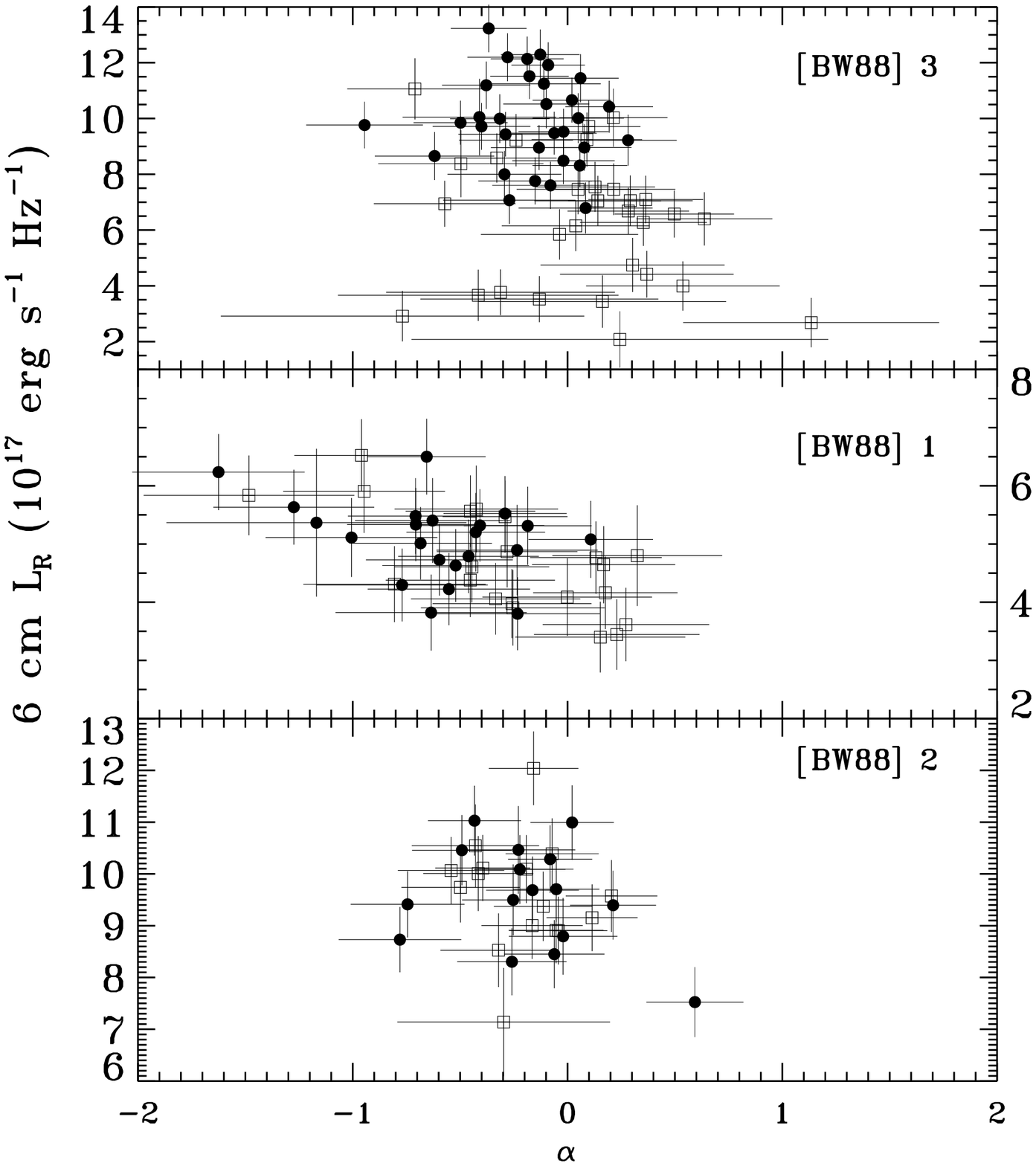}
\includegraphics[scale=0.35,width=3in]{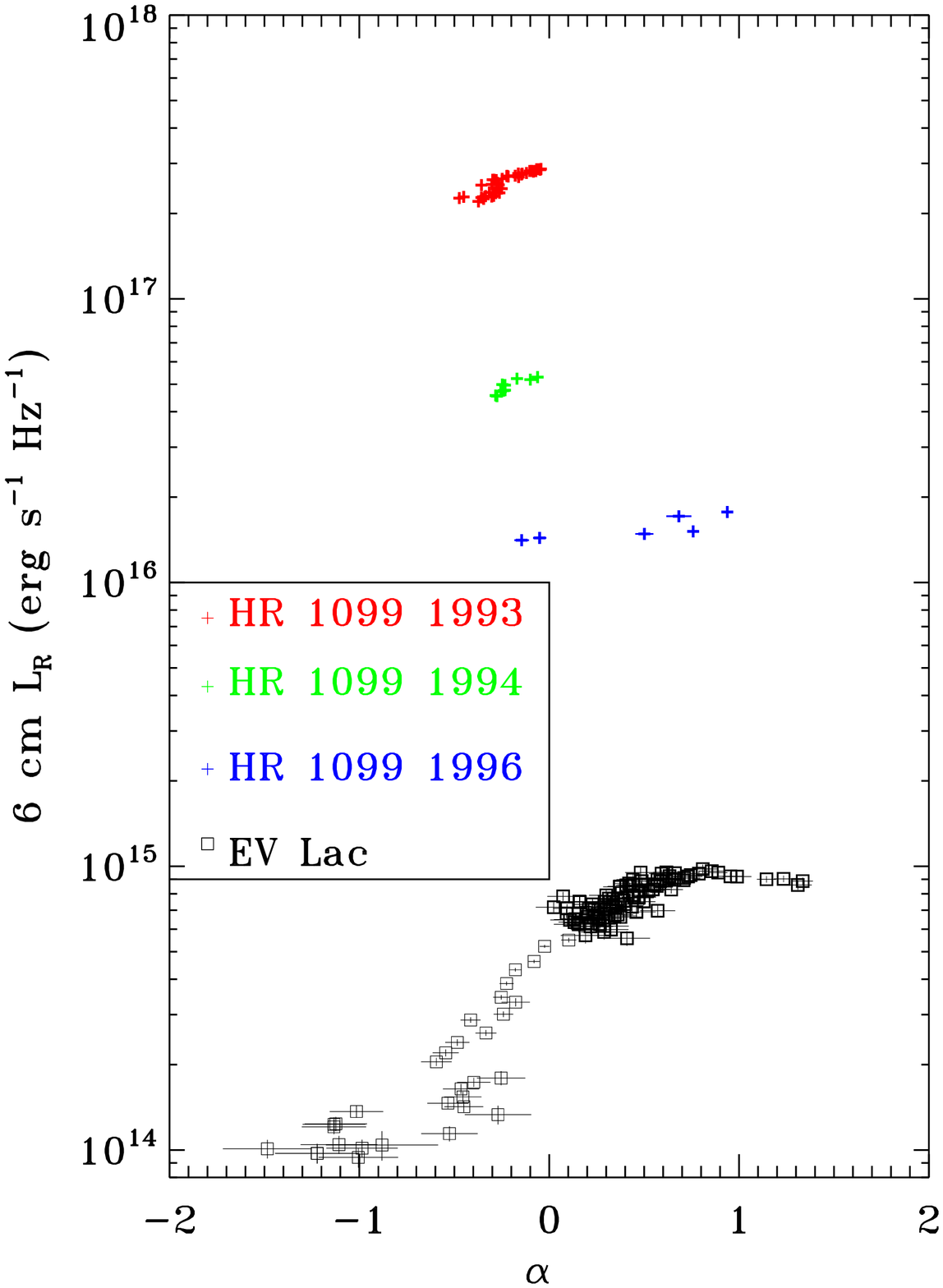}
\caption{{\bf (left)} Plot of 6 cm radio luminosity against 6--3.6 cm spectral index, for
three objects for which changes in spectral index could be measured. 
Open squares indicate data points from the first day of observations, while the
filled circles denote data from the second observation.  
The two apparent cluster members, [BW88]~1 and [BW88]~3, display a
statistically significant anti-correlation between luminosity and spectral index, while the probable
AGN ([BW88]~2) shows no correlation.
{\bf (right)} Plot displaying 6 cm radio luminosity against 6--3.6 cm spectral index
for the decays of several radio flares from nearby active stars. 
Data for decays of radio flares
from HR~1099 are taken from \citet{osten2004}; data for decay of radio flare from EV~Lac is taken
from \citet{osten2005}.  The radio luminosities are positively correlated with spectral index
over a wide range of luminosity.
\label{fig:alpha_flux}}
\end{center}
\end{figure}

The panels in Figure~\ref{fig:radiolc} show the temporal variation of spectral index for three
objects with large enough flux density at both frequencies to permit determination
of changes in the spectral index.
[BW88]~3 is the only source where a large variation in flux density appears to be occurring during our
observation.  
The light curve of [BW88]~1 also shows some short time-scale variability predominantly
at 6 cm which also reveals itself as large negative values of $\alpha$.  
The behavior of radio luminosity versus spectral index is displayed in Figure~\ref{fig:alpha_flux}
for the three objects in which spectral index changes can be discerned.
The anti-correlation between luminosity and spectral index is statistically significant
for sources [BW88]~3 and [BW88]~1 at the $>$99.9\% level.
In addition, we investigated the behavior between luminosity and spectral index for
each day separately, as the light curves indicate differing amounts of
variability on each day.  For [BW88]~3, the anti-correlation is significant only
at the 87\% level on the first day (March 6), despite the obvious large-scale flux variations, and
is not statistically significant on the second day (March 8).
For [BW88]~1, the anti-correlation was significant at $>$99.7\% and 90\%, respectively,
for the two days.  For [BW88]~2, there was no correlation between radio luminosity and spectral index.

\subsection{X-ray Fluxes}
Our X-ray spectral fitting procedure is discussed in detail in 
Paper~II and follows closely our
previous work \citep{wolk2006,wolk2008}. Spectra were fitted to 
about 100 sources with over 30 counts using several
models for emission spectra from thermal, diffuse gas including that
of \citet{rs1977}, \citet{apec} and mekal \citep{mewe1985}.
In summary, in Paper~II we
find in the LkH$\alpha$~101 cluster that the
mean temperature of the coronae is about 2.5 keV, with the  typical
range between 800 eV and 5 keV.   There were however 25 sources with
temperatures $>$ 10 keV which were excluded from this analysis due to
the low collecting area of the $Chandra$ HRMA.  Typical \nh\ values
ranged from 0-4 $\times 10^{22}$ cm$^{-2}$ with an outlier rejected
mean of about 7.8  $\times 10^{21}$ cm$^{-2}$.

The spectroscopically measured absorbed
fluxes range from 4.73$\times10^{-13}$ to $4.89\times10^{-15}$ erg
cm$^{-2}$ sec\minusone.  Assuming a distance of 700~pc and correcting 
the measured X-ray absorption this becomes a luminosity range of about 
$log~L_X= 29.75-31.64$ erg sec\minusone.  
We compute the on-axis X-ray detection limits for a 3$\times 10^{7}$K plasma (kT$\sim$
2.5 keV)
with absorbing column density $\log N_{H}=$21.9 near the center of the field,
and require 5 photons for a detection.  The minimum absorbed flux
is 6.2$\times$10$^{-16}$ erg cm$^{-2}$ s$^{-1}$, or $log~L_X \sim 28.5$ erg sec\minusone  
for $d=700$ pc.  This value is intended to give a rough
sense of the sensitivity of the observation; variable absorption has a
strong influence on the luminosity limit, as does off-axis distance. 
About twice the noted flux is required 4\amin\ off-axis. 

\subsection {X-ray Variability \label{xrvary}}
X-ray variability of the sources in this field is discussed more
thoroughly in the companion paper. 
Two independent analysis methods were used. First was a Bayesian block
analysis which is common in the X-ray literature 
\citep[BBs;][]{scargle1998,getman2005,wolk2006}.
For the Bayesian analysis we used two different priors,
one prior was set to find variability at about 95\%  confidence and the
second was set to 99.9\% confidence.  We also analyized the lightcurves
following the method of \citet[][GL-vary]{gl1992}. 
This method also uses Bayesian statistics and evaluates a
large number of possible break points from the prediction of
constancy. The advantage of the GL-vary method is that it handles
data gaps and changes to the effective area well.
It also calculates the odds that a source was constant.
Thus, the analysis returns the probability that the source
was variable and an estimate of the constant intervals within the
observing window.  All analyses were repeated on the data from the
individual observations. 
We compare the results of Bayesian analysis with those of the more standard
KS test by 
using the {\it Chandra} Orion Ultradeep Project (COUP) dataset \citep{getman2005} as an example:  the BB algorithm found  973 variables,  
95\% of these were considered variable by the KS test at more than 99.9 \% probability. 

\begin{figure}[htbp]
\includegraphics[scale=0.5]{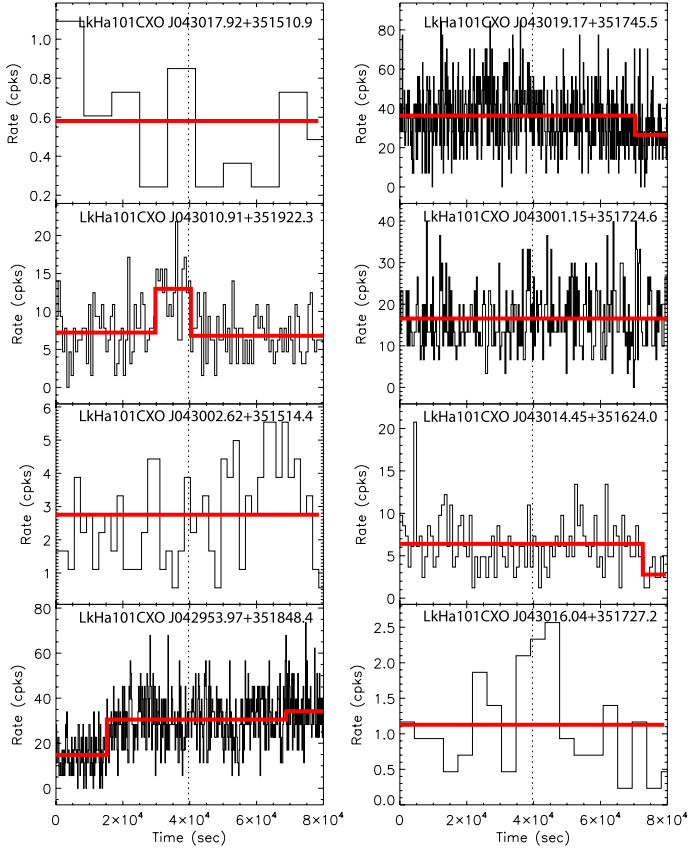}
\caption{
$Chandra$ light curves of radio-detected sources.  The
  vertical line represents the break between the first observations, 
(ObsID 5429; 6~Mar 2005) and the second (ObsID 5428; 8~Mar 2005).  
The time histogram is binned so that the average bin
contains 5 counts. Hence they range from about 100 seconds to about
9~ks in length.  The red lines indicate intervals of constant flux as
determined by Bayesian analysis set to detect variability at $>$ 95\%
confidence. IAU designators for source names are used. 
Count rate is counts per kilosecond.
\label{fig:xraylc}}
\end{figure}

For the purposes of this paper we consider only the eight X-ray sources with
radio counterparts.  The lightcurves for these sources are shown in Figure~\ref{fig:xraylc}.
Half of the sources show variability at $>95\%$ confidence based on
the GL-vary method (sources LkHa101CXO J043019.2+351745, LkHa101CXO J043010.9+351922,
LkHa101CXO J043002.6+351514 and LkHa101CXO J042954.0+351848).  However no
flaring or variations greather than a factor of
two were seen.   In fact, all of the sources seemed very stable between the two
observations.   The fact that no flares were seen is not particularly
remarkable.  The product of the observing time and the number of
sources is about 640 ksec.  This is about the mean time between
macroscopic flares seen in young clusters 
\citep[c.f.][]{wolk2005,wolk2006,caramazza2007}.

\section{Results}
These deep X-ray and radio observations probe multi-wavelength
signatures of magnetic activity 
from the stars in the LkH$\alpha$ 101 star forming region.
This paper focuses on
the radio sources and associated multi-wavelength data and what we can learn 
about magnetic activity in these young stellar objects.
There are three facets to the study described in
this paper: the nature of the radio emitting sources, the variability of magnetic activity,
and correlations between radio and X-rays, which we describe in the next sections. 

\subsection{On the Nature of the Radio-Emitting Cluster Members}
Out of the 16 radio sources considered here, 8 are X-ray sources and 7 have Spitzer counterparts.
There is a large overlap between radio-detected X-ray sources and radio-detected
Spitzer sources: all of the IRAC sources are X-ray sources.
The combination of X-ray and infrared detections bolsters support for
these radio-emitting objects to be cluster members.
Their close distance to the cluster center, and X-ray and infrared properties also make
these objects consistent with being cluster members.
In the following subsections, we provide a brief summary of the characteristics determined for each radio-emitting
cluster member.
The multi-wavelength identifications are summarized in Table~\ref{tbl:idtbl}.

\subsubsection{LkHa101VLA J043017.90+351510.0 = [BW88]~1}
[BW88]~1 was detected in the radio and X-ray maps, but does not have a detection in
the Spitzer maps.  There are no 2MASS point sources within 17" of the position of [BW88]~1 \citep{2mass}.
It is located in a region of high nebulosity, hence the infrared detection threshold requires 
about an additional magnitude compared to clear regions.  Nonetheless, based on the upper limits in the Spitzer bands,
if this object is a bona-fide member of the cluster, then it may be close to the substellar limit.

This is a bright and persistent radio source, detected on 5 out of 5 occasions (Table~\ref{tbl:radiotbl1}).  Our simultaneous
observations at 3.6 and 6 cm reveal a negative spectral index, suggesting nonthermal emission.  This source 
shows no variability at
radio wavelengths over the $\sim$20 years spanning the archival data and our data, 
but does appear to be variable at $>$ 3 $\sigma$ during the two days of our observations at C band (but not X band), and 
on shorter timescales at both wavelengths.  The X-ray observations reveal no evidence of variability.
The multi-frequency radio observations show an anti-correlation between radio flux density and spectral index.
This object also displays a value of radio to X-ray luminosity more than 10 times that of other
cluster members (\S 4.3).
If this is a substellar object, the 
anomalously high ratio of its radio to X-ray luminosity when compared to the values
found for the other stellar objects would fit with the few nearby field very low mass stars which
display large radio luminosities relative to their X-ray luminosities \citep{berger2006}.

\subsubsection{LkHa101VLA J043019.14+351745.6 = [BW88]~3}
[BW88]~3 has been detected in the radio, X-ray, and Spitzer maps, and is a Class~II object
based on its IR colors.  
\citet{herbig2004} identified this source as a K0V star.

At radio wavelengths it is a persistent source, having been detected
on 6/6 occasions, with a negative radio spectral index.  
The source shows large flux variations 
over the longest timescales probed here, and is variable over the two days in the most recent epoch
at both X \& C bands; it also displays variability on the shortest timescales, with a factor of 6 contrast between
the highest/lowest flux densities. 
The large-scale short-term radio variability observed at 3.6 and 6 cm could be radio flares or
rotational modulation.
If rotational modulation, this would imply a likely period of $\approx$ 1--2 days and could arise from
either geometrical effects or, alternatively, some form of directivity of the emission, similar to
what has been seen in radio observations of the pre-main sequence star AB~Doradus \citep{lim1994}.  
Such an interpretation might imply that the emitting particles are much
higher in energy than usually assumed for gyrosynchrotron emission.
The radio observations on short timescales show an anti-correlation between radio flux
density and spectral index, similar to that seen in [BW88]~1.  There is statistical evidence
for X-ray variability but no obvious X-ray flaring.  

\subsubsection{LkHa101VLA J043010.87+351922.4 = [BW88]~4}
[BW88]~4 was detected at radio, X-ray, and IR wavelengths, and is a Class~III object 
based on its IR colors. 
\citet{bw} argue that this is an obscured B dwarf, noting a small H II region around it.

At radio wavelengths [BW88]~4 has only been detected on 2 occasions out of 5, and only at C band, with a
factor of $\approx$ 7 difference between the recorded flux densities.  While this demonstrates
large long-term radio variability, there are no significant variations over the shorter timescales
probed by the two days of our observations.
We have no constraints on
radio spectral index.  
Multiple Bayesian blocks indicate X-ray variability on the first day.

\subsubsection{LkHa101VLA J043001.15+351724.6 = [BW88]~6}
[BW88]~6 has been detected at radio, X-ray, and IR wavelengths, and is a Class~III object
based on its IR colors.
It has been identified by \citet{herbig2004} as a K7 dwarf with Lithium.
At radio wavelengths, it has been detected on 3 out of 6 occasions, with factor of $>$10
variability in two C-band detections.  In our radio observations the object was detected
only by summing all data from our two days' integration at C band, so we have no constraints
on shorter timescale variability, nor any constraint on spectral index.  There is also no
evidence for X-ray variability.

\subsubsection{LkHa101VLA J043002.64+351514.9 = [BW88]~9}
[BW88]~9 was detected in previous radio observations, and is detected in X-ray and Spitzer maps,
with a Class~II designation based on its IR colors.  There is no spectral type for this object.
It is only an intermittent radio source,
having been detected on 2 out of 6 occasions at radio wavelengths.  There is statistical evidence
for X-ray variability but no large-scale flares.  

\subsubsection{LkHa101VLA J043014.43+351624.1 = LkH$\alpha$101}
LkH$\alpha$101 is the source of the reflection nebula and H~II region seen in optical/IR and radio images, and is also 
the source of a strong stellar wind.  It has been detected at radio, X-ray, and IR wavelengths.
The IR class is unknown due to the large amount of nebulosity,  but observations at other wavelengths
show the presence of a secondary as well as a face-on disk \citep{tuthill2001,tuthill2002}.
The radio spectral index is large and positive, consistent with being a wind source.
There is apparent variability between the X and C band flux densities measured on the two days, but this is
likely not intrinsic to the source.  There is no evidence for X-ray variability.

\subsubsection{LkHa101VLA J042953.98+351848.2}
This is a newly detected radio source, which has X-ray and IR counterparts, and based on
its IR colors it is a class~III object.  No spectral type is known.  
This object was only detected at C band due to the wider full-width half power beam size,
and was only detected on the second day of radio observations.
It thus indicates radio variability between the two days as well as short-term
radio variability.  There is statistical evidence for X-ray variability 
and multiple Bayesian blocks on the first day of X-ray observations, but no
variability on the second day when it is detected at radio wavelengths.

\subsubsection{LkHa101VLA J043016.04+351726.9}
This is another newly detected radio source, which has X-ray and IR counterparts,
and is a class~III object based on its IR colors.  It has a spectral type
of dM2 according to \citet{herbig2004} and a flat or slightly negative radio spectral index.
There is radio variability at C band over the two days with factor of two variations, but none at X band.
There is no evidence for X-ray variability. 

\subsection{Variability and Magnetic Activity}
\subsubsection{Timescales of Magnetic Activity at Radio Wavelengths}
There are three different timescales which we have explored
for evidence of radio variability, and we find that roughly half the
objects examined on each timescale are variable.
Extragalactic objects could be radio variable
due to AGN activity or interstellar scintillation, while cluster members could have
variable radio emission due to magnetic activity.
On the longest timescales of 20+ years, 4/9 sources with detections at 6 cm had more than
10$\sigma$ variations between the maximum recorded flux and the minimum or upper
limit of recorded flux.  Two of these sources appear to be cluster members.
On the scale of days, 5/9 sources at 3.6 cm
and 6/14 sources at 6 cm varied by more than 3$\sigma$, 
with two of the 5 variable sources at 3.6 cm being cluster members and
5 of the 6 variable sources at 6 cm being cluster members based on multi-wavelength
identifications.
On the smallest
timescales of 1200--2400 s, 2/5 sources at 3.6 cm and 4/7 sources at 6 cm
had a $>$ 95\% probability of being variable; two of the variable sources at 3.6 cm
and 6 cm
are cluster members.

If we restrict ourselves to objects with counterparts at X-ray or IR wavelengths
and thus probable cluster members,
then 2/5 cluster members show evidence of variability on the longest timescales,
1/3 and 4/5 show day-to-day variations of more than 3 $\sigma$ at 3.6 and 6 cm,
respectively, and 2/2 and 2/3 cluster members had a $>$95\% probability of being variable
on the shortest timescales at 3.6 and 6 cm, respectively.
The small number of objects prevents
a conclusive statistical analysis of whether cluster members are
more likely to be radio variable than the background extragalactic objects.
Four of the 8 sources detected at both radio and X-ray wavelengths showed
evidence of being X-ray variable, despite lacking large-scale flares or 
factors of 2 or more variability.
On the shortest timescales, radio variability appears to be a more common feature
of cluster members than is X-ray variability.

\subsubsection{Rapid Radio Variability} 
The radio observations on the shortest timescales reveal evidence of statistically
significant flux changes 
in the majority of cluster members for which
such measurements are possible.
While the radio variability between the two days could be due to the contribution of different
magnetic structures, the rapid radio variability is likely due to magnetic reconnection.
The timescales for the rapid radio variability span from $>$ t$_{\rm bin}$ of 1200--2400 seconds
to $\sim$ 11 hours, the duration of the large flux density enhancement seen in [BW88]~3.
If we interpret these rapid variations as magnetic reconnection flaring, 
the timescale of the rapid radio variability
argues for a low density environment.  
The Coulomb deflection times for particles of energy $E_{\rm kev}$ in a magnetic trap with ambient density
$n_{e}$ cm$^{-3}$,\\
\begin{equation}
t_{\rm d} = 0.8 \frac{E_{\rm kev}^{3/2}}{(n_{e}/10^{8})} \frac{20}{\log \Lambda} \;\;\; s
\end{equation}
where $\Lambda$ is the Coulomb logarithm (evaluated at the mean cluster temperature of 2.5 keV),
constrains $t_{\rm d}$ to be $\gtrsim$1200--2400 seconds (time binning of the
shortest variations) and $\lesssim$ 11 hours (the timescale for large flux density variations).
For a 20 keV electron, this will occur for a range roughly $10^{5} \lesssim n_{e} (cm^{-3}) \lesssim 5\times10^{6}$,
and for a 200 keV electron $5\times 10^{6} \lesssim n_{e} (cm^{-3}) \lesssim 2\times10^{8}$.
We note that quantitative analyses of the
time-dependent radio flux spectrum have been done for RS CVn binaries
\citep{rscvns} but not, to our knowledge, for the more complicated
magnetic field geometries around young stars \citep{donati2007}.
Such low densities are to be contrasted with the ambient electron density in
X-ray-emitting coronal material, which is generally much higher \citep[$n_{e} \ge$10$^{9}$ cm$^{-3}$;][]{jardine2006}.

\subsubsection{Rapid Radio Variability and Lack of Circular Polarization}
The constraints on circular polarization from the maps made of all the data, described in
\S 2.1, can be applied to the measured flux densities at 3.6 and 6 cm to constrain the
amount of circular polarization.  At 3.6 cm, the minimum and maximum values of flux density 
imply 3 $\sigma$ limits on the percent of circularly polarized flux  $V/I$ of
$<$2\% and $<$ 29\%, respectively, while at 6 cm the corresponding limits
are $<$ 2\% and $<$ 31\%, respectively.  


The high level of radio variability implies a rapid re-arrangement of magnetic fields, but the
lack of circular polarization potentially places a constraint on the spatial distribution of magnetic fields.  
Orientation can affect the
observed amount of circular polarization for large-scale magnetic structures (edge-on configurations
reveal contributions from both hemispheres, leading to a net cancellation).  Alternatively, 
a magnetic configuration with many small-scale structures would also lead to a small net 
value of circular polarization.
Even when 
circular polarization is detected in young stellar objects, it is usually at a low level
\citep[2--4\% in quiescence, up to $\sim$16\% during flares;][]{white1992}. 
Our upper limits on circular polarization are thus only weakly constraining.

\subsubsection{Anti-correlation Between Flux Density and Spectral Index}
There are two cluster members showing rapid radio variability which have strong enough
detections to allow investigation of temporal changes in spectral index ([BW88]~1 and [BW88]~3).
The spectral indices are generally flat or negative, $\alpha \lesssim$0.5, indicating 
nonthermal emission from an inhomogeneous source.
The anti-correlation between 6 cm radio luminosity and spectral index
(\S3.4)
is opposite to that seen from solar flares and
active stars \citep{benz1977,mutel1987}.
This anti-correlation has
also been observed previously in young stellar radio sources:
\citet{felli1993} noted that some of the radio sources in Orion 
had spectral index changes associated with flux density changes on timescales of weeks,
with $\alpha$ becoming more negative when the flux density
increases. 
During the
decays of radio flares from nearby active stars 
the spectral index and flux are positively correlated.
The right panel of Figure~\ref{fig:alpha_flux} displays the 6 cm radio luminosity versus 6--3.6 cm spectral index
for 4 well-observed radio flares from the active binary HR~1099 and the dMe flare star EV~Lac
\citep{osten2004,osten2005}.  The positive correlations between radio luminosity and spectral index for these
flares are significant at $>$99.5\% confidence.  Thus the observed trends in the 
two young stellar objects discussed here reveal a divergent behavior from that exhibited 
in active stars.

\begin{figure}[htbp]
\includegraphics[scale=0.3,angle=90]{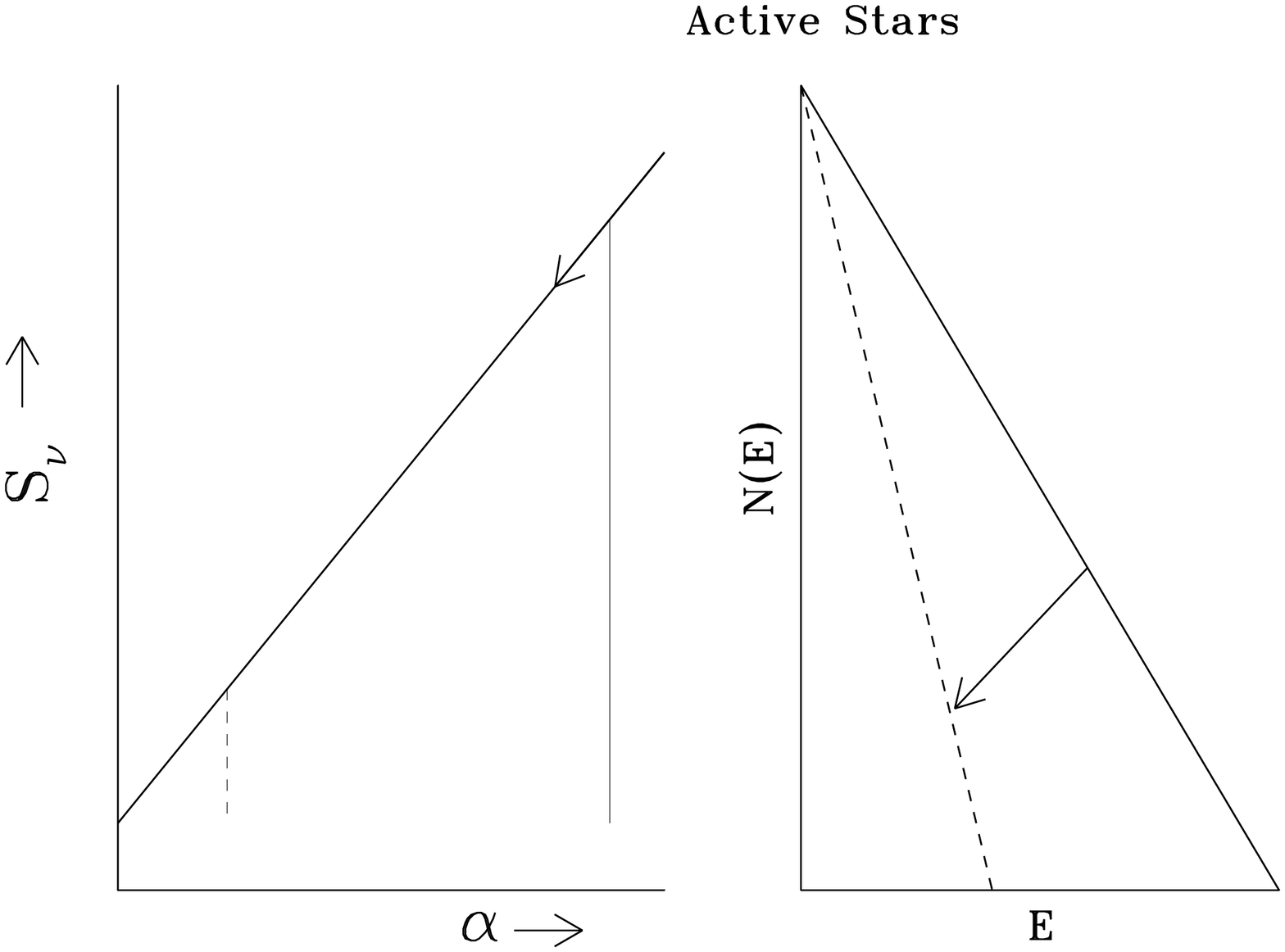}
\includegraphics[scale=0.3,angle=90]{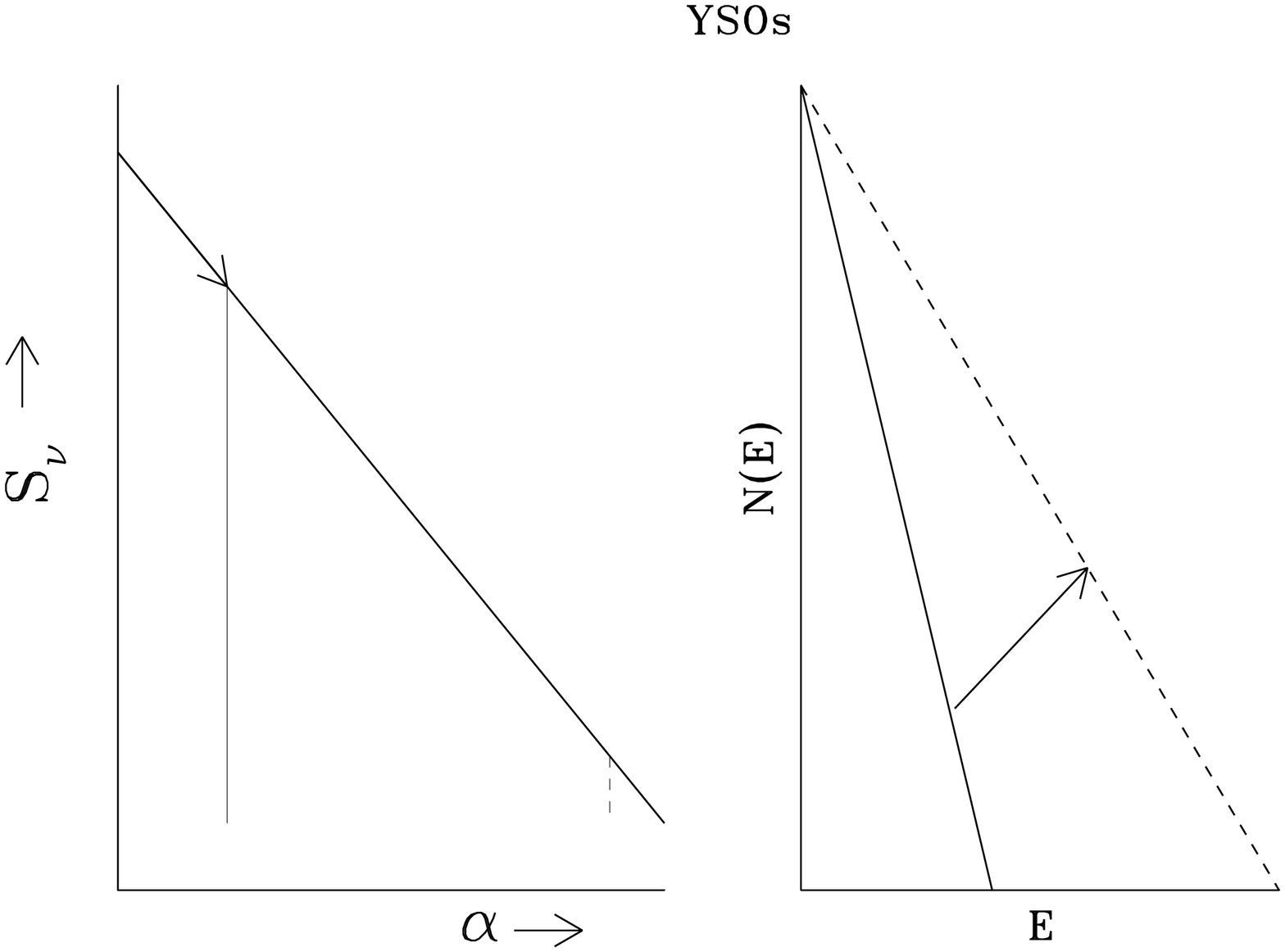}
\caption{ {\bf (left)} Schematic relationship between flux density and spectral index
observed in the decay phase of flares on active stars. 
Solid vertical line in S$_{\nu}$ vs. $\alpha$ plot indicates $S_{\nu}$, $\alpha$ measurements at a time early in the flare
decay, while dashed vertial line indicates $S_{\nu}$, $\alpha$ measurements at a later
time in the flare decay. 
Solid slanted line in $N(E)$ vs. $E$ figure indicates relative
distribution of accelerated particles at the time early in the flare decay, while
slanted dashed line indicates that for later in the flare decay, based on the change of $\alpha$
under optically thin conditions.
The interpretation of the decrease
in $\alpha$ is a a softening of the accelerated electron population during the flare decay.
{\bf (right)} Schematic relationship between flux density and spectral index for two
young stellar objects seen here and in \citet{felli1993}.  Solid and dashed lines are the same as
for the left panels.
The hardening of the accelerated
electron spectrum from early in the flare decay to later times
implies sporadic or near-continuous acceleration of electrons to
repopulate the distribution.  
\label{fig:schm}
}
\end{figure}

A detailed modelling of the flux density and spectral index and their temporal variations
is beyond the scope of this paper.  Nevertheless, we can examine the observed trends to
infer constraints on the evolution of the population of accelerated particles.
For radio flares from active stars the positive correlation 
between flux density and spectral index
is considered to be a consequence of spectral evolution
of optically thick emission during the rise and peak phases of the flare, returning to optically thin
values during the flare decay.
Under optically thin conditions, the spectral index is a function of only the
power-law index of the accelerated electrons.  
Specifically, 
$\alpha \sim 1.2-0.9 \delta$ for $2\leq \delta \le 7$ \citep{dulk1985}. 
Thus, for active stars, the positive
correlation between flux density and spectral index means that at the peak of the flare
when the flux density is maximum, $\alpha$ is large, corresponding to small values of $\delta$
which implies a hard distribution.
At further points in the flare decay, the flux density has declined, with consequent smaller
values of $\alpha$ and larger values of $\delta$.  This temporal evolution is towards
a softer accelerated electron distribution, such as can happen when the most energetic electrons
are depopulated from e.g. radiative losses in a high B field region (left panel, Figure~\ref{fig:schm}).

The anti-correlation between the flux density and spectral index seen in two cluster
members suggests a different scenario than what is observed for active stars.  For [BW88]~3
the temporal evolution of the flux density on both days looks like the decay phase of two
flares, yet when the flux density is maximum, the spectral index is at its smallest.
The temporal variations of [BW88]~1 are not ordered in the same way, yet produce the same anti-correlation.
A small spectral index, under the same optically thin conditions as assumed above, implies a large
value of $\delta$, while the large $\alpha$ corresponding to smaller values of flux density
implies a {\it smaller} value of $\delta$ (right panel, Figure~\ref{fig:schm}).  This evolution of the accelerated particle spectrum
is from soft to hard, and is difficult to envision under standard conditions of energy losses. 
This implies that the hardest energy electrons are being repopulated.  For [BW88]~3 this
timescale is several hours, whereas for [BW88]~1 it is on the order of the time binning, $\sim$1200 seconds.
The conditions present in [BW88]~1 thus argue for a rapid but sporadic particle acceleration, while those
in [BW88]~3 argue for nearly continuous particle acceleration during the transient events seen.

\subsection{Radio/X-ray Correlations \label{rx}}
\begin{figure}
\begin{center}
\includegraphics[scale=0.8]{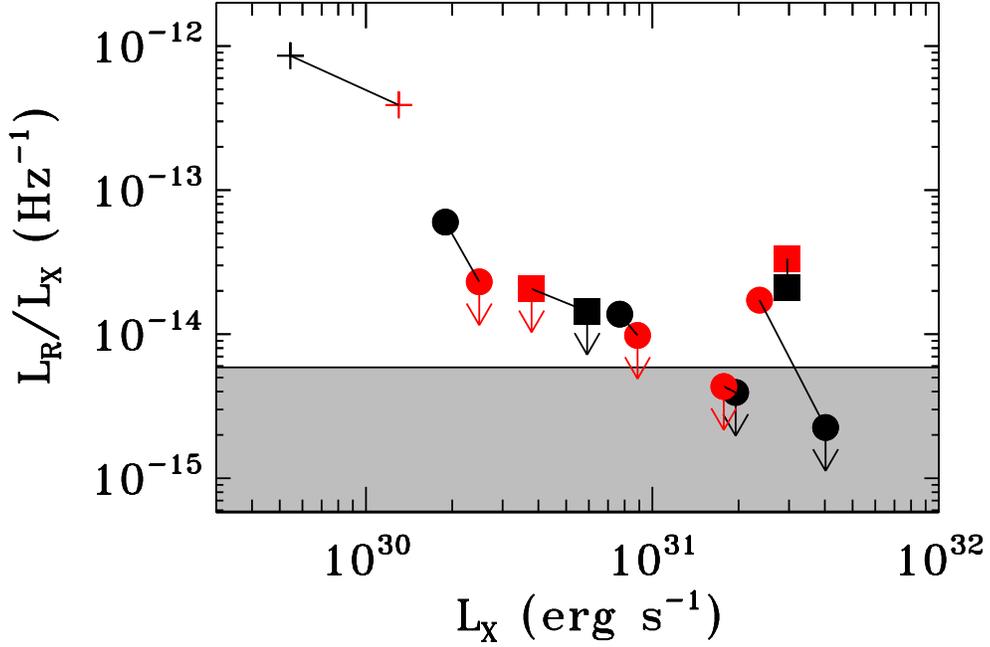}
\caption{ 
Ratio of radio to X-ray luminosities for radio sources considered in the
present study.
Black points indicate measurements on the first day (03/06), red points 
measurements obtained from the second day (03/08), and a line connects measurements
of the same object. Class~II objects are denoted by filled squares, while Class~III objects
have a filled circle for their data points.  
Data for the one object not
detected with Spitzer are delineated by plus symbols. 
Down arrows indicate upper limits.
The shaded curve
indicates the region of $L_{X}-L_{R}$ space occupied by the GB
relationship, with the order of magnitude uncertainty in the relationship, and
constant appropriate to cTTs.  
It appears likely that these objects lie outside this relationship.
\label{lxlr}}
\end{center}
\end{figure}

We investigated radio and X-ray luminosity correlations for objects
detected at both radio and X-ray wavelengths, utilizing the simultaneously
obtained measurements from our project.  Figure~\ref{lxlr} displays the results
graphically, with X-ray and radio luminosity measurements taken from each of the two
days of observations.  We used a distance of 700 pc \citep{herbig2004},
and the flux density measurements at 6 cm.
The dotted lines connecting the pairs of points indicate the
classification of the source using IRAC colors:  there were 4 objects whose
colors indicate photospheric levels, two whose colors were consistent with Class II
sources, and one object not detected in the IRAC bandpass.  
We exclude the central object LkH$\alpha$101 from consideration, due to
its early spectral type and the likelihood that the radio emission arises
from an ionized wind.
Two sources have only
radio upper limits due to their nondetection at the current epoch, compared with
detections at earlier epochs at levels up to five times the current upper limits.
Also plotted is the expected range of
X-ray and radio luminosities based on the observational relationship 
found by \citet[][hereafter, GB]{gb1993}.
One might only expect the GB relationship to hold in wTTs but
since we see that 1 cTTs displays nonthermal radio emission (see discusion in \S
3.2) we include the Class~II
objects as well.

Figure~\ref{lxlr} indicates that the radio and X-ray luminosities displayed by
these objects is not consistent with that implied by the GB relationship.
For the Class II and III objects, there is considerable scatter in the measured values,
with none of the radio and X-ray detections being consistent with the GB
relationship (which would result in a ratio L$_{R}$/L$_{X}$ of 5.9--59$\times$10$^{-16}$ Hz$^{-1}$
for wTTs).  
The prevalence of these objects above the 
GB relationship in the $L_{R}$-L$_{X}$ plot indicates that they are more radio-luminous
at a given X-ray luminosity than the GB relationship would imply.
The sensitivity of our radio observations leads to a 5$\sigma$ constraint on 
radio luminosity of $\sim$6$\times$10$^{16}$ erg s$^{-1}$ Hz$^{-1}$ for a source close to
the phase center; based on arguments in \S 3.5 we estimated an X-ray sensitivity of $\sim$3$\times$10$^{28}$
erg s$^{-1}$ for an on-axis source at 700 pc distance.  The X-ray sensitivity is capable of
exploring a larger region of $L_{X}$-$L_{R}$ parameter space than is the current radio sensitivity.
Note that two of our upper limits to radio luminosity could be consistent with the GB L$_{X}$-L$_{R}$
relationship with radio observations $\approx$ 2--3 times more sensitive.

Magnetic fields are complicit in plasma heating which produces X-ray
emission; magnetic fields also accelerate electrons and give rise to radio emission.
It is natural, then, to assume that a relationship might hold between two observable
sources of radiation whose formations require the presence of magnetic fields.
\citet{gb1993} argued that the two quantities can be roughly linearly related through conversion of 
coronal energy into plasma heating and X-ray radiation on the one hand, and particle acceleration
and radio emission on the other hand.  Another way the two quantities may be linearly related is if 
the radio- and X-ray-emitting volumes are co-spatial.  But this places a stringent
constraint on the X-ray-emitting plasma, because the radio radiation must be able to escape
and be detected.  Using parameters from the X-ray spectral fits to the integrated data
for radio- and X-ray detected sources, and assuming a spherically symmetric corona with
stellar radii of $\approx$ 2R$_{\odot}$, roughly appropriate for stars
of about 1 Myr,
we estimate the free-free optical depth of the X-ray emitting material at radio wavelengths
\citep[equation 11.2.2 in ][]{benz2002}
and find that the X-ray emitting material would be optically thick at radio wavelengths.
This rules out co-spatial formation of X-ray and radio emission.
To produce detectable quantities of radio and X-ray emission requires the presence of large scale
magnetic fields and/or large amounts of magnetic flux.  It is entirely possible that
with a complex field geometry, X-ray and radio-emitting regions are physically distinct
from each other, with separate energy reservoirs.  If this is the case, no relationship between
the two emergent intensities is to be expected.  
 
Figure~\ref{lxlr} displays the ratio of the luminosities against X-ray luminosity.
There is a statistically significant anti-correlation between this
luminosity ratio and the X-ray luminosity;
the discrepancy between the observed luminosity ratio and 
that expected from the GB relationship grows as the X-ray luminosity decreases.
It is likely that [BW88]~1, the most extreme
object exhibiting this relationship, is a very low mass star or brown dwarf (see discussion in \S 3.1).
The luminosity ratio relates the efficiency of particle acceleration (through the production of
nonthermal radio emission) to that of plasma heating (through the production of X-ray emission).
Although the GB relationship implies a nearly constant amount of particle acceleration relative to plasma
heating, the results for the young stellar objects around LkH$\alpha$101 
indicate that this parameter depends on the amount of plasma heating, and
thus implies a decoupling of particle acceleration from plasma heating.

\begin{figure}[htbp]
\includegraphics[scale=0.7,angle=90]{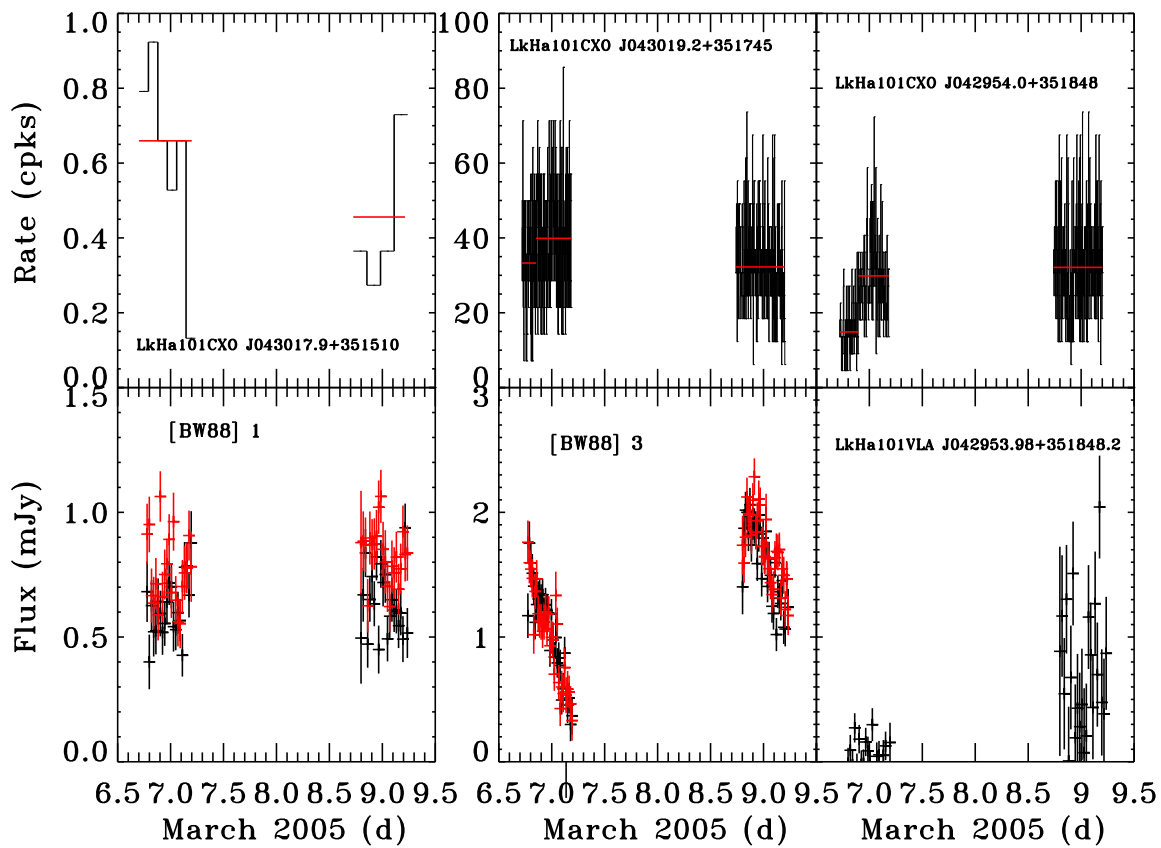}
\caption{
X-ray and radio light curves for three objects with radio fluxes
high enough to examine short time-scale radio variability.  
Top panels: Chandra light curves as in Figure~\ref{fig:xraylc},
bottom panels: radio light curves as in Figure~\ref{fig:radiolc}.
There is no apparent correlation between radio and X-ray variability 
in these objects.
\label{fig:xrr}}
\end{figure}

The lack of correlation between radio and X-ray measurements can also be seen
in the comparison of short time-scale variability:  
The amount of radio variability between
the two days of observations is larger than the corresponding range in 
X-ray flux in the sample of objects detected at both wavelengths.  
Out of 8 sources detected at both X-ray and radio wavelengths, variability at both
wavelengths 
is seen in only two cases ([BW88]~3, LkHa101VLA J0429540.+351848).
Yet, two objects appeared to be undergoing short time-scale variability ([BW88]~3, [BW88]~1), while no X-ray
flares were seen.
The comparison of X-ray and radio variability is shown in Figure~\ref{fig:xrr} for 
the three objects having significant radio detections, to probe variability in both.
Variability appears to be a greater factor in the production of radio emission
than it is in X-ray emission.

\section{Discussion}
There has been a paucity of simultaneous radio and X-ray observations
of young stars, despite the large potential for learning about 
magnetic fields and interrelation of plasma heating and particle acceleration.
The small number of cases in hand suggest a different scenario from that
gleaned from studies of the Sun and nearby active stars.
Simultaneous multi-wavelength observations of HD~283447 by
\citet{feigelson1994}
revealed a radio flare but no X-ray or
chromospheric variability.  They concluded that the regions of gyrosynchrotron
emission were largely decoupled from the regions where plasma heating was occurring.
\citet{bower2003} serendipitously observed a giant outburst
at mm- wavelengths during the COUP observation;
an intense X-ray flare on the same star began 3 days prior to
the observed radio flare and lasted through the duration of the radio
flare.
\citet{gagne2004} found a star, DoAr 21, in the $\rho$ Opch cloud core to be in the
the decay phase of an X-ray flare with
stable emission at 6 cm.
These examples are opposite to the classic ``Neupert effect'' and suggestive
of a different flare scenario.  
Time-averaged correlations also show a different
behavior in young stellar objects.
\citet{gagne2004} found no evidence that X-ray and radio
luminosities are correlated as expected in the GB relationship for a small sample of TTs 
in the $\rho$ Ophiucus cloud. 
Observations of Class~I sources in the Coronet cluster \citep{forbrich2006,forbrich2007}
similarly show no clear
correlations between radio and X-ray emission as should appear in the GB relationship
if particle acceleration and plasma heating arise from a common energy reservoir. 
Our finding of a decoupling between radio and
X-ray emission in both a time-averaged and 
time variable sense supports the idea that the structures giving rise to
the two emissions in young stellar objects are physically or energetically distinct.

We find as a common theme that there is a different scenario for magnetic activity in the young
stellar objects around LkH$\alpha$101 than is usually seen in the Sun and active stars.
This is revealed by a comparison between radio flux versus spectral index behavior, which suggests for 
the young stellar objects studied here rapid particle acceleration and injection.  The 
interpretation of the dependence of radio to X-ray luminosity ratio on X-ray luminosity 
is that the efficiency of particle acceleration relative to plasma heating depends on the amount
of plasma heating, rather than being relatively independent as for active stars and the Sun.
We investigated the coronal structures which could be giving rise to the radio and X-ray emission,
and find several lines of evidence that such a decoupling is occurring.  The X-ray emitting loops are optically
thick to radio emission at the wavelengths observed, so the detected amounts of radio and X-ray emission
can not be coming from co-spatial loops. 
The isolation of radio- and X-ray emitting material is also consistent with one interpretation of the lack of
expected roughly linear correlation between X-ray and radio luminosities.
The observed rapid radio variability implies a low density environment, another disconnect between the
spatial location 
of radio emitting structures and X-ray emitting structures, as 
 inferred X-ray densities are much higher.

We can also constrain the size scales of X-ray and radio emission.  Since the observed X-ray luminosity is
proportional to the volume emission measure (VEM), a function of the coronal electron density and size scale, for a spherically symmetric corona we have \\
\begin{equation}
VEM \propto \int n_{e}^{2} dV \sim 4\pi R_{\star}^{2} \int n_{e}^{2} ds
\end{equation}
where R$_{\star}$ is the stellar radius, which we take for YSOs to be roughly 2 (see discussion in \S 4.3).
Using the observed X-ray luminosities and derived column emission measures in \S 4.3, combined with coronal
density estimates of n$_{e}$ between 0.6--2.5 $\times$10$^{10}$ cm$^{-3}$ \citep{jardine2006}, implies size scales
for the X-ray emission of $l_{X}/R_{\star}$ between 0.002 and 2, and thus relatively compact.  
For gyrosynchrotron emission appearing at the observed radio frequencies, valid for harmonics of the gyrofrequency 
$\nu=s\nu_{B}$ for $s$ between 10 and 100 with $\nu_{B}$=2.8$\times$10$^{6}$B Hz, the inferred magnetic field strengths
in the region of radio emission are between 17 and 300 G.  If we assume that this emission originates from a dipole field geometry,
and the surface magnetic fields are in the range 1--3 kG \citep{cmj2007}, then the scale size of the radio emission
$l_{R}$ is between 1.6 and 6 R$_{\star}$.  This shows that the radio emission appears to be physically distinct
from the X-ray emission.  This conclusion has also been reached by \citet{feigelson1994} and
\citet{massi2006} for the case of HD~283447.

If there is a large-scale distribution of magnetic fields giving rise to the radio
emission, the lack of circular polarization constrains the orientation of the emission to be
largely edge-on, while a magnetic configuration composed of small-scale fields would also explain
a null detection of circular polarization in radio emission.
Recent magnetic field measurements in TTs show that strong and complex magnetic field geometries
are present, which can potentially interfere with X-ray generation \citep{cmj2007}
and probably radio generation as well.
\citet{jardine2006} noted from modelling the X-ray emission of T Tauri stars that more complex
field geometries were associated with more compact, denser coronae.
Perhaps the complex field
configuration and possible coupling of disk to star changes the emergent radiations
associated with the magnetic fields in such a way as to break the $L_{x}-L_{r}$ relation
and any variability correlations.

\section{Conclusions}
We find that nonthermal radio emission
(based on flat/negative spectral indices) can be produced even in stars with infrared evidence for disks.
In fact, we see large-scale radio emission variability from one Class~II object ([BW88]~3),
as well as a possibly substellar object ([BW88]~1).
Thus radio emission can be used as a diagnostic of magnetic activity in young stellar objects both with 
and without disks.  
Using this observational tool, we find a variety of behaviors indicating a different kind of magnetic activity 
from that seen on active stars and the Sun:
anti-correlation between radio luminosity and radio spectral index, 
lack of expected correlation between radio and X-ray luminosities, lack of correlated 
variability between radio and X-ray emission, and anti-correlation between the ratio of radio
to X-ray luminosity and X-ray luminosity.
The anti-correlation between radio flux and spectral 
index points to a gyrosynchrotron emission mechanism which requires a different
evolution of field strength, number density of accelerated electrons, and distribution
compared to what is seen on nearby active stars and the Sun.
The multi-wavelength behaviors suggest a decoupling between particle acceleration and plasma heating, 
in both time-averaged and time-variable trends.
The lack of expected correlation between radio and X-ray luminosities can be
interpreted as due to their production in different magnetic regions, or out of different
energy reservoirs.  There are a few calculations which support the former:
the different inferred electron densities for X-ray emitting material versus that required
to explain the rapid variability of radio emission (X-ray emission arising from denser regions);
the calculated optical depth to radio emission of the X-ray-emitting plasma (showing that
the X-ray emitting plasma would be optically thick and therefore not likely to escape
and be detected at radio wavelengths); and the lack of
correspondence between estimated X-ray size scales and radio size scales, which suggest for
simple configurations that the X-ray emission is more compact than the radio emission.

The primary finding of our investigation is that the magnetic activity signatures (radio, X-ray) and their
correlations from young stellar objects in the LkH$\alpha$101 cluster show different trends
from those established for nearby active stars.  
Our results are new in that we probe a region of variability space occupied by few other
multi-wavelength observations of young stellar objects: long-duration, multi-frequency radio coverage,
with simultaneous multi-wavelength observations.
We conclude that we are likely seeing evidence of a new phenomenon in these highly active
young stars.
Our conclusions are tempered by the small number of objects in our analysis, which is a result of
the asymmetry between percentages of radio and X-ray emission in cluster members.
We are expanding our multi-wavelength investigation in magnetic activity from young stars
with current and future observations of clusters spanning a range in ages.
The simultaneous multi-wavelength coverage in this campaign was vital to establishing the
disconnect between these magnetic activity signatures, and points out the 
parameter space available for future such studies. 
Future observations with the Expanded Very Large Array (EVLA), with roughly a factor of
ten increase in sensitivity to radio emission, will be able to expand upon the findings presented here.

\acknowledgements
We especially thank Tyler Bourke,  Rob Gutermuth and Brad Spitzbart --- co-authors
on Paper~II upon whose results we are building.
Thanks to Jan Forbrich for an advance copy of Forbrich \e 2007.
Thanks also to the referee, Marc Gagn\'{e}, for constructive comments
on improving this paper.
This publication makes use of data products from the Two Micron All Sky Survey,
which is a joint project of the University of Massachusetts and the Infrared
Processing and Analysis Center, funded by the National Aeronautics and Space
Administration and the National Science Foundation.
Support for this work was provided by NASA through Hubble Fellowship grant \# HF-01189.01 awarded
by the Space Telescope Science Institute, which is operated by the Association of Universities for
Research in Astronomy, Inc. for NASA, under contract NAS5-26555.  
The CXC guest investigator program supported this work through grant GO5-6018X.
SJW was supported by NASA contract NAS8-03060.
This represents the results of VLA program S60872.


\clearpage

\begin{deluxetable}{lcccccccc}
\tablewidth{0pt}
\tablenum{1}
\rotate
\setlength{\tabcolsep}{0.03in}
\tabletypesize{\scriptsize}
\tablecolumns{9}
\tablecaption{Long-term Behavior of Radio Sources in the Field around
  LkH $\alpha$101 \label{tbl:radiotbl1}\tablenotemark{1}}
\tablehead{\colhead{ID} & 
\colhead{April 1985} & \colhead{January 1986} & \colhead{April 1986} & \colhead{May 1986} & \colhead{October 1991}
 & \colhead{March 2005} & \colhead{March 2005 }  & \colhead{$\alpha$\tablenotemark{a}} \\
 \colhead{} &  \colhead{6 cm} & \colhead{6 cm} & \colhead{6 cm} & \colhead{6 cm} & \colhead{3.6 cm} & \colhead{3.6 cm} & \colhead{6 cm} & \colhead{}  }
\startdata
LkHa101VLA J043017.90+351510.0    &       {\it 0.7$\pm$0.1} &     {\it 1.1$\pm$0.1}  &   {\it  1.0$\pm$0.2} &      NI         &    {\it  0.63$\pm$0.07 }  &0.629$\pm$0.013 &  0.794$\pm$0.016&   -0.43$\pm$0.05 \\
LkHa101VLA J043026.04+351538.2    &    {\it    2.0$\pm$0.1} &  {\it    2.2$\pm$0.1}  & {\it  2.1$\pm$0.2} &     {\it  1.4$\pm$0.1 } &   {\it   1.65$\pm$0.12}   &1.428$\pm$0.016 &  1.694$\pm$0.019&   -0.32$\pm$0.03 \\
LkHa101VLA J043019.14+351745.6   &     {\it  2.9$\pm$0.1} &  {\it     1.2$\pm$0.1}  &  {\it   2.2$\pm$0.1} &     {\it  4.2$\pm$0.2}  &   {\it   1.12$\pm$0.11 }  &         1.276$\pm$0.014 &  1.347$\pm$0.018 &  -0.10$\pm$0.03 \\
LkHa101VLA J043010.87+351922.4   &   {\it    1.3$\pm$0.1} &  {\it    $<$0.1 }      & {\it     $<$0.4}      &  {\it    $<$0.2 }      &     OOF          & OOF  &      0.222$\pm$0.023   &-- \\
LkHa101VLA J043003.74+351827.6   &  {\it    0.9$\pm$0.1} &   {\it   2.0$\pm$0.1 } &     OOF        &      NI         &   {\it   0.46$\pm$0.07 }       & 0.616$\pm$0.018 &  0.832$\pm$0.021 &  -0.55$\pm$0.07 \\
LkHa101VLA J043001.15+351724.6   &   {\it    0.5$\pm$0.1} &    {\it   $<$0.1  }     &  {\it    $<$0.4}  &       {\it    $<$0.2   }    &    {\it  0.33$\pm$0.08}  & OOF            &  0.056$\pm$0.016        &  -- \\
LkHa101VLA J042956.40+351553.4   &   {\it    1.2$\pm$0.1} &   {\it    1.8$\pm$0.1 } &  {\it   $<$0.4\tablenotemark{b}} &      {\it 0.7$\pm$0.1}  &     {\it 0.22$\pm$0.06 } & OOF            &  0.578$\pm$0.022 &  -- \\
LkHa101VLA J042956.78+351527.1   &   {\it    0.5$\pm$0.1} &    {\it   0.8$\pm$0.1}  &  {\it    1.1$\pm$0.2\tablenotemark{b}}  &     {\it      0.6$\pm$0.1}  &   {\it   0.50$\pm$0.06}  & OOF            &  1.249$\pm$0.025 &  -- \\
LkHa101VLA J043002.64+351514.9   &  {\it     $<$0.2}&       {\it     0.8$\pm$0.1}  &  {\it    $<$0.4}  &        {\it   $<$0.2 }      &  {\it    0.81$\pm$0.10}  & $<$0.072        &  $<$0.059        &  -- \\
LkHa101VLA J042958.00+351602.9  &      $<$0.3 &            $<$0.2         &     $<$0.3    &          NI         &     {\it 0.82$\pm$0.17}        & OOF            &  0.117$\pm$0.022 &  -- \\
LkHa101VLA J043008.77+351625.8   &      $<$0.3 &            0.22$\pm$0.06         &     OOF    &          NI         &     {\it 1.05$\pm$0.14 }       & $<$0.036        &  $<$0.036       &   -- \\
LkHa101VLA J043010.98+351437.6   &      $<$0.3 &            $<$0.2         &     OOF    &          NI         &     {\it 0.29$\pm$0.05 }       &   0.058$\pm$0.014&   $<$0.054    &      $>$0.2 \\
LkHa101VLA J043014.43+351624.1    &NI &NI    & NI                  &  NI         &  NI                   &    38.134$\pm$0.011         & 24.643$\pm$0.014 & 0.805$\pm$0.001 \\
LkHa101VLA J043014.46+351527.5    &      $<$0.3 &            $<$0.2         &     OOF    &          NI         &     {\it 0.38$\pm$0.06  } &$<$0.036    &      0.067$\pm$0.015 & $<$-1 \\
LkHa101VLA J043016.53+351710.8   &     $<$0.3  &            $<$0.2         &     OOF    &          NI         &     {\it 0.55$\pm$0.10 }        &$<$0.036    &      $<$0.045 &         -- \\
LkHa101VLA J043016.98+351642.4    &      $<$0.3  &            $<$0.2         &     OOF    &          NI         &     {\it 0.65$\pm$0.14 }   &        0.043$\pm$0.011   &       0.066$\pm$0.015     &   -0.8$\pm$0.6 \\
LkHa101VLA J043017.34+351647.6     &      $<$0.3  &            $<$0.2         &     OOF    &          NI         &     {\it 0.54$\pm$0.08 }        &         0.103$\pm$0.012   &       0.063$\pm$0.015     &     0.9$\pm$0.4 \\
LkHa101VLA J043009.74+351502.5     &      $<$0.3  &            $<$0.2        &     OOF    &          NI         &     $<$0.18           & 0.138$\pm$0.013 &   0.149$\pm$0.018  & -0.1$\pm$0.3 \\
LkHa101VLA J043002.85+351709.8     &      $<$0.3  &            $<$0.2         &     $<$0.3    &          NI         &     $<$0.3           & 0.129$\pm$0.016 &   0.171$\pm$0.018  & -0.5$\pm$0.3 \\
LkHa101VLA J043004.01+351817.0     &      0.36$\pm$0.12 &            $<$0.2         &     OOF    &          NI         &     $<$0.3           & $<$0.054        &   0.227$\pm$0.020  & $<$-3 \\
LkHa101VLA J042953.98+351848.2     &      $<$0.3  &            $<$0.2         &     OOF    &          NI         &     OOF           & OOF            &   0.353$\pm$0.027  & -- \\
LkHa101VLA J043016.04+351726.9     &      $<$0.3  &            $<$0.2         &     OOF    &          NI         &     $<$0.2           & 0.103$\pm$0.012 &   0.124$\pm$0.016  & -0.4$\pm$0.3 \\
LkHa101VLA J043024.78+351757.7      &      $<$0.4  &            $<$0.2         &     OOF    &          NI         &     OOF           & 0.323$\pm$0.016 &   0.223$\pm$0.021  & 0.6$\pm$0.2 \\
\enddata
\tablenotetext{1}{Primary beam-corrected peak intensities and 1$\sigma$ rms values are listed for
detections; 3 $\sigma$ upper limits are given otherwise.  Numbers from the four epochs described in BW or that of SO
for sources listed in 
either BW or SO are taken from those papers and are delineated in italics.  
Measurements of other
sources from these datasets are the purview
of this paper.  ``OOF" means the position of the source was out of the nominal image field; ``NI" 
means no information on this source at this epoch is available.
A ``--" in the column for spectral index means that flux information at one or both
frequencies was not available, usually because the object was outside the field of view at one frequency 
or undetected at both frequencies.}
\tablenotetext{a}{$\alpha$ is the slope from 6--3.6 cm, defined as S$_{\nu} \propto \nu^{\alpha}$.}
\tablenotetext{b}{These two numbers are switched from what is reported in Becker \& White, based on our examination
of the archival data.}
\end{deluxetable}

\begin{deluxetable}{clcccl}
\tablewidth{0pt}
\tablenum{2}
\rotate
\setlength{\tabcolsep}{0.03in}
\tabletypesize{\scriptsize}
\tablecolumns{6}
\tablecaption{Multi-wavelength cross-identification of sources \label{tbl:idtbl}}
\tablehead{\colhead{Radio ID} & \colhead{other Radio ID \tablenotemark{a}} & \colhead{X-ray ID}  & \colhead{IRAC counterpart} & \colhead{IR Class\tablenotemark{b}} & \colhead{spectral type}
}
\startdata
LkHa101VLA J043017.90+351510.0 &[BW88]~1, [SO98]~12 &  LkHa101CXO J043017.9+351510  & & & \\
LkHa101VLA J043026.04+351538.2 &      			 [BW88]~2, [SO98]~14 &  & & &probable AGN(see \S4.1.2)\\
LkHa101VLA J043019.14+351745.6 &[BW88]~3,[SO98]~13 & LkHa101CXO J043019.2+351745 &LkHa101SST J043019.16+351745.5&  2 & K0: \citep{herbig2004}\\
LkHa101VLA J043010.87+351922.4 & [BW88]~4 & LkHa101CXO J043010.9+351922 &LkHa101SST J043010.89+351922.6& 3 &BV (BW88)\\
LkHa101VLA J043003.74+351827.6 & 			 [BW88]~5,[SO98]~4&  && &probable AGN (see \S4.1.2)\\
LkHa101VLA J043001.15+351724.6 & [BW88]~6, [SO98]~16 & LkHa101CXO J043001.1+351724 &LkHa101SST J043001.13+351724.8 & 3 & K7 \citep{herbig2004}\\
LkHa101VLA J042956.40+351553.4 & 			 [BW88]~7, [SO98]~15 & && & probable AGN (see \S4.1.2)\\
LkHa101VLA J042956.78+351527.1 & 			[BW88]~8, [SO98]~1 & & &&\\
LkHa101VLA J043002.64+351514.9 & [BW88]~9, [SO98]~3 &LkHa101CXO J043002.6+351514 &LkHa101SST J043002.62+351514.4&  2  &\\
LkHa101VLA J042958.00+351602.9 &				 SO~2&&	&  &\\
LkHa101VLA J043008.77+351625.8 & 			 SO~5& & &&\\
LkHa101VLA J043010.98+351437.6 & 			 SO~6 && &&\\
LkHa101VLA J043014.43+351624.1 & LkH$\alpha$101, SO~7 & LkHa101CXO J043014.4+351624 &LkHa101SST J043014.44+351624.0  & 100 &early B; see discussion in \citet{herbig2004}\\
LkHa101VLA J043014.46+351527.5 & 			SO~8& & &&\\
LkHa101VLA J043016.53+351710.8 & 			SO~9 &&&&\\
LkHa101VLA J043016.98+351642.4 & 			SO~10 &&& &\\
LkHa101VLA J043017.34+351647.6 &				SO~11&& & &\\
LkHa101VLA J043009.74+351502.5 &				&    & & &\\
LkHa101VLA J043002.85+351709.8 &				&    & & &\\
LkHa101VLA J043004.01+351817.0 & 			&     & &&\\
LkHa101VLA J042953.98+351848.2 & & LkHa101CXO J042954.0+351848       &LkHa101SST J042953.97+351848.6& 3& \\
LkHa101VLA J043016.04+351726.9 & &LkHa101CXO J043016.0+351727       &LkHa101SST J043016.04+351727.2& 3 & M2 \citep{herbig2004}\\
LkHa101VLA J043024.78+351757.7 &			       &     & &\\
\enddata
\tablenotetext{a}{Key to IDs: BW=\citet{bw},SO=\citet{so}}
\tablenotetext{b}{Spitzer classification code: 2=Class II, 3=Class III, 100=unknown}
\end{deluxetable}

\begin{deluxetable}{lccccccc}
\tablewidth{0pt}
\tablenum{3}
\rotate
\setlength{\tabcolsep}{0.03in}
\tabletypesize{\scriptsize}
\tablecolumns{8}
\tablecaption{Two-Day Behavior of Radio Sources in the Field around LkH$\alpha$101\label{tbl:radiotbl2}\tablenotemark{1}}
\tablehead{\colhead{ID} & \colhead{$\theta$ } & 
\colhead{03/06} & \colhead{03/06} & \colhead{03/08} & \colhead{03/08} & \colhead{$\alpha_{1}$\tablenotemark{a}} & \colhead{$\alpha_{2}$\tablenotemark{a}}  \\
\colhead{} & \colhead{('')} & \colhead{3.6 cm} & \colhead{6 cm} & \colhead{3.6 cm} & \colhead{6 cm} & \colhead{} & \colhead{}
}
\startdata
LkHa101VLA J043017.90+351510.0 &  84 &   0.614$\pm$0.018 & 0.760$\pm$0.021  &  0.639$\pm$0.018  & 0.831$\pm$0.021 &  -0.39$\pm$0.07  &  -0.48$\pm$0.07 \\
LkHa101VLA J043026.04+351538.2 &  149 &   1.343$\pm$0.022 & 1.678$\pm$0.025  &  1.518$\pm$0.022  & 1.708$\pm$0.024 &  -0.41$\pm$0.04  &  -0.22$\pm$0.04 \\
LkHa101VLA J043019.14+351745.6 &  102 &   0.9894$\pm$0.019 & 0.992$\pm$0.035  &  1.580$\pm$0.019  & 1.672$\pm$0.023 &  0$\pm$0.6   &  -0.10$\pm$0.03 \\
LkHa101VLA J043010.87+351922.4 &  185 &   OOF            & 0.214$\pm$0.030  &  OOF             & 0.218$\pm$0.031        &  --            &  -- \\
LkHa101VLA J043003.74+351827.6 &  181&   0.636$\pm$0.025 & 0.850$\pm$0.028  &  0.590$\pm$0.026  & 0.813$\pm$0.028 &  -0.53$\pm$0.09  &  -0.6$\pm$0.1 \\
LkHa101VLA J043001.15+351724.6 &  174 &   OOF            & $<$0.072         &  OOF             & $<$0.072        &  --            \\
LkHa101VLA J042956.40+351553.4 &  222 &   OOF            & 0.625$\pm$0.030  &  OOF             & 0.562$\pm$0.029 &  --            &  -- \\
LkHa101VLA J042956.78+351527.1 &  222 &   OOF            & 1.178$\pm$0.033  &  OOF             & 1.310$\pm$0.032 &  --            &  -- \\
LkHa101VLA J043002.64+351514.9 &  159 &   $<$0.081        & $<$0.081         &  $<$0.069         & $<$0.081        &  --            &  -- \\
LkHa101VLA J043014.43+351624 &  2 &  36.168$\pm$0.015 & 23.099$\pm$0.019 &  37.788$\pm$0.015 & 23.196$\pm$0.018&  0.827$\pm$0.002 &  0.900$\pm$0.002 \\
LkHa101VLA J043009.74+351502.5 & 98 &   0.084$\pm$0.017 & 0.113$\pm$0.022  &  0.183$\pm$0.019  & 0.148$\pm$0.022 &  -0.5$\pm$0.5    &  0.4$\pm$0.3 \\
LkHa101VLA J043002.85+351709.8 &  149 &   0.139$\pm$0.021 & 0.167$\pm$0.044  &  0.124$\pm$0.022  & 0.149$\pm$0.024 &  -0.3$\pm$0.3    &  -0.4$\pm$0.4 \\
LkHa101VLA J043004.01+351817.0 &  171&    $<$0.072        & 0.171$\pm$0.024  &  $<$0.072         & 0.129$\pm$0.025 &  $<$-1.6         &  $<$-1 \\
LkHa101VLA J042953.98+351848.2 &  289&    OOF            & $<$0.150         &  OOF             & 0.724$\pm$0.036 &  --            &  -- \\
LkHa101VLA J043016.04+351726.9  &  68 &   0.135$\pm$0.017 & 0.188$\pm$0.021  &  0.125$\pm$0.017  & 0.093$\pm$0.020        &  -0.6$\pm$0.3    &  0.5$\pm$0.4 \\
LkHa101VLA J043024.78+351757.7  &  159   &   0.405$\pm$0.023 & 0.232$\pm$0.029  &  0.282$\pm$0.022  & 0.201$\pm$0.026 &  1.2$\pm$0.2     &  0.6$\pm$0.3 \\
\enddata
\tablenotetext{1}{Primary beam-corrected peak intensities and 1$\sigma$ rms values are listed for detections; 3$\sigma$ upper limits given
for nondetections.  ``OOF'' means that the position of the source was out of the nomial image field.
A ``--" means that flux information at one (or both) frequencies was not available, usually because
the object was outside the field of view at one frequency or undetected at both frequencies.}
\tablenotetext{a}{$\alpha_{1}$ is the slope from 6--3.6 cm on 6 March, $\alpha_{2}$ the slope from 6--3.6 cm on 8 March.}
\end{deluxetable}

\begin{deluxetable}{lccccccc}
\tablewidth{0pt}
\tablenum{4}
\rotate
\setlength{\tabcolsep}{0.03in}
\tabletypesize{\scriptsize}
\tablecolumns{7}
\tablecaption{Variability Statistics \label{tbl:varstat}}
\tablehead{ \colhead{ID} & \colhead{Long Term Var.\tablenotemark{a} } & \colhead{2-day Var.\tablenotemark{b} } 
& \colhead{2-day Var.\tablenotemark{b}} & \colhead{Short Term Var.\tablenotemark{c} }
& \colhead{Short Term Var.\tablenotemark{c} }
& \colhead{P$_{\rm XR}$\tablenotemark{d}} \\
\colhead{} & {6 cm} & \colhead{3.6 cm} & \colhead{6 cm}  & \colhead{3.6 cm } & \colhead{6 cm }  & \colhead{(\%)}
}
\startdata
LkHa101VLA J043017.90+351510.0 & N & N & Y & Y &  Y  & 60 \\
LkHa101VLA J043026.04+351538.2 & N & Y &N & N & N  & - \\
LkHa101VLA J043019.14+351745.6 & Y & Y&Y & Y & Y &  97 \\
LkHa101VLA J043010.87+351922.4 & Y & - & N & - & - & 95 \\
LkHa101VLA J043003.74+351827.6 & Y & N & N & N & Y  & - \\
LkHa101VLA J043001.15+351724.6 & N & - & - & - & - & 17 \\
LkHa101VLA J042956.40+351553.4 & Y & - & N &   - & N & - \\
LkHa101VLA J042956.78+351527.1 & N & - & Y &  -& Y & -  \\
LkHa101VLA J043002.64+351514.9 & N & - & - & - & - & 99 \\
LkHa101VLA J043014.43+351624.1 & - & Y & Y & -\tablenotemark{e} & -\tablenotemark{e} & 72 \\
LkHa101VLA J043009.74+351502.5 &- & Y & N & - & - & - \\
LkHa101VLA J043002.85+351709.8 & - & N &N & - & - & - \\
LkHa101VLA J043004.01+351817.0 & - & - & N & - & - & - \\
LkHa101VLA J042953.98+351848.2 & - & - & Y & - & N  & $>$99.9 \\
LkHa101VLA J043016.04+351726.9 & - & N &Y &-  & - & 82 \\
LkHa101VLA J043024.78+351757.7 & - & Y &N & N  & - & - \\
\enddata
\tablenotetext{a}{
See \S~\ref{longt} for explanation of how long-term radio variability was assessed.}
\tablenotetext{b}{Variability between two days of radio observations based on $>$3$\sigma$ difference in flux density;
see \S~\ref{mediumt} for explanation.}
\tablenotetext{c}{
Short timescale radio variability; see \S~\ref{shortt} for details.}
\tablenotetext{d}{Probability that the source is X-ray variable; see \S~\ref{xrvary}.}
\tablenotetext{e}{Embedded in bright nebulosity; see text for details.}
\end{deluxetable}

\clearpage

\end{document}